\documentclass[
amsmath,amssymb,eshowkeys,nofootinbib,linebumbers,
prd,
twocolumn
]{revtex4-2}

\usepackage{color}
\usepackage{comment}
\usepackage{graphicx}
\usepackage{dcolumn}
\usepackage{bm}

\usepackage[colorlinks=true,linkcolor=red,urlcolor=blue,citecolor=blue]{hyperref}

\usepackage{soul, xcolor}
\soulregister\cite7
\soulregister\ref7
\soulregister\eqref7

\definecolor{skyblue}{RGB}{135,206,235}

\newcommand{\bchi}{\bar{\chi}}
\newcommand{\hchi}{\hat{\chi}}
\newcommand{\bhchi}{\bar{\hchi}}
\newcommand{\cA}{\mathcal{A}}

\newcommand{\hcA}{ \hat{{\mathcal A}}}
\newcommand{\fF}{\mathsf{F}}
\newcommand{\bfF}{\bar{\fF}}
\newcommand{\bOm}{\bar{\Omega}}
\newcommand{\tomega}{\kappa}
\newcommand{\absangparam}{\ell}
\newcommand{\angparam}{m_\ell}
\newcommand{\azimuthal}{m_j}
\newcommand{\pmass}{m}
\newcommand{\dbar}{\lower0.1ex\hbox{$\mathchar'26$}\mkern-12mu d}

\newcommand{\nn}{\nonumber}

\newcommand{\p}{\partial}

\newcommand{\mOm}{\mathsf \Omega }
\newcommand{\hmOm}{\hat{\mOm}}

\newcommand{\whsF}{{\widehat{\mathsf{F}}}}
\newcommand{\cF}{{\mathcal F}}
\newcommand{\hsF}{{\hat{\mathsf{F}}}}
\newcommand{\hcF}{{\hat{\cF}}}
\newcommand{\bscF}{ \boldsymbol{{\mathcal F}}}

\newcommand{\cL}{{\mathcal L}}

\newcommand{\hcD}{\hat{{\mathcal D}}}
\newcommand{\cG}{{\mathcal G}}

\newcommand{\whcG}{\widehat{{\mathcal G}}}
\newcommand{\mJ}{\mathrm J}
\newcommand{\whmJ}{\widehat{\mJ}}

\newcommand{\Tr}{\mathop{\rm Tr}\nolimits}

\newcommand{\etat}{\tilde{\eta}}
\newcommand{\etab}{\bar{\eta}}

\newcommand{\chib}{\bar{\chi}}
\newcommand{\tchi}{\tilde{\chi}}

\newcommand{\bM}{\bar{M}}

\newcommand{\cS}{{\mathcal S}}

\newcommand{\mT}{\mathsf T}

\newcommand{\mV}{\mathsf V}

\newcommand{\etach}{\check{\eta}}

\begin{document}

\title{Revisiting the fermionic quasibound states around Schwarzschild black holes with improved analytic spectrum}%

\author{Guang-Shang Chen$^{1,2}$}
\author{Cheng-Bo Yang$^{1,2}$}
\author{Shou-Shan Bao$^3$}
\email{ssbao@sdu.edu.cn}
\author{Yong Tang$^{4,6,7}$}
\email{tangy@ucas.ac.cn}
\author{Yue-Liang Wu$^{1,4,5,6}$}
\email{ylwu@itp.ac.cn}
\affiliation{%
\begin{footnotesize}
 $^1$ CAS Key Laboratory of Theoretical Physics, Institute of Theoretical Physics, 
 Chinese Academy of Sciences, Beijing 100190, China \\
 $^2$ School of Physical Sciences, University of Chinese Academy of Sciences, No. 19A Yuquan Road, Beijing 100049, China \\
 $^3$ Institute of Frontier and Interdisciplinary Science,
 Key Laboratory of Particle Physics and Particle Irradiation (MOE), Shandong University, Qingdao 266237, China \\
 $^4$ International Centre for Theoretical Physics Asia-Pacific (ICTP-AP), UCAS, Beijing 100190, China \\
 $^5$ Taiji Laboratory for Gravitational Wave Universe (Beijing/Hangzhou), University of Chinese Academy of Sciences (UCAS), Beijing 100049, China \\
 $^6$ School of Fundamental Physics and Mathematical Sciences, Hangzhou Institute for Advanced Study, UCAS, Hangzhou 310024, China \\
 $^7$ School of Astronomy and Space Sciences, University of Chinese Academy of Sciences (UCAS), Beijing 100049, China
\end{footnotesize}
}%

\date{\today}

\begin{abstract}
Black holes have long served as a testing ground for probing theories of gravity and quantum mechanics. Notably, fundamental fields in the neighborhood of black holes exhibit rich phenomena that could yield astrophysically observable signatures. However, exploring these structures typically requires computationally intensive numerical calculations. In this work, the dynamics of a massive Dirac field outside a Schwarzschild black hole are revisited. We propose a novel matching scheme that enables the analytical solution of the coupled first-order Dirac equation, as opposed to the conventional second-order approach. This method yields a compact and unified analytical expression for the energy spectrum, which shows improved agreement with the numerical results. The improvement comes from the higher-order correction of the angular parameter that was ignored previously.
\end{abstract}

\maketitle

    \section{Introduction}
    \label{sec_intro}
    
	The \textit{no-hair} conjecture~\cite{Israel:1967za, Carter:1971zc} states that classical black holes in static equilibrium are characterized by only a few parameters such as mass $M$, angular momentum $J$ and charge $Q$. However, in reality, they are perturbed. Even isolated from the macroscopic objects, distortion from quantum fields exists~\cite{Konoplya:2011qq}. Bosonic fields may undergo quasiresonance or quasibound states~\cite{Ternov:1978gq, Grain:2007gn, Hod:2015goa, Konoplya:2004wg}, exhibit wigs~\cite{Barranco:2012qs} and even form metastable clouds~\cite{Arvanitaki:2009fg, Brito:2015oca}. Recently, an improvement in the matching technique~\cite{Starobinsky:1973aij, Detweiler:1980uk} to analyze the Klein-Gordon equation for light bosons was analyzed thoroughly in~\cite{Bao:2022hew}, which addressed the long-standing discrepancy of a factor of two with numeric results. 
	
	In contrast to the extensive works on bosons, there are only a few about fermions in curved spacetime. A key distinguishing feature for fermions is the absence of superradiance due to the Pauli exclusion principle. Unruh first shows this feature for the massless fermion by employing a representation in which the Dirac equation is separable in the Kerr background~\cite{Unruh:1973bda}. The separability has soon been generalized to the massive field~\cite{Chandrasekhar:1976ap, Unruh:1976fm} and reformulated with the help of the notion of the generalized total angular momentum operator~\cite{Carter:1979fe}. Based on the separated form of the Dirac equation in Kerr spacetime, the absence of superradiance for the massive fermion has been proven~\cite{Gueven:1977dq, Martellini:1977qf, Lee:1977gk, Iyer:1978du}. More details can be found in Chandrasekhar's book~\cite{Chandrasekhar:1985kt} and references therein. We are intrigued by the possibility of neutrinos orbiting around a primordial black hole~\cite{Villanueva-Domingo:2021spv}. Besides the scattering of the massless neutrino~\cite{Brill:1957fx, Chandrasekhar:1977kf} and the Dirac quasinormal modes~\cite{Cho:2003qe}, neutrinos with a tiny mass can form quasibound systems with the black hole. Fermions, captured or emitted during evaporation, take chances to fill up all the quasilevels and create a dense Fermi sea~\cite{Hartman:2009qu, Coutant:2018bom}. The condensed system, also called a dark atom, may be a possible candidate for dark matter~\cite{Dokuchaev:2014vda}. 
	
	For details of such structures, it is essential to delve into the spectrum of quasibound states. The spectrum of the Dirac field was first calculated using both matching techniques~\cite{Ternov:1980st} and standard perturbation methods~\cite{Gaina:1988nf}. These works focus on one component governed by a Klein-Gordon-like second-order equation obtained by eliminating the other one. Toward a more comprehensive understanding of fermions coupled to gravity, a detailed analysis of the full Dirac equations is necessary. Many efforts thus far have been devoted to numerical investigations~\cite{Doran:2005vm, Dolan:2009kj, Dolan:2015eua, Huang:2017nho, Konoplya:2017tvu}. However, simulations require high precision, as quantum mechanics in curved spacetime combines simultaneously quantities at atomic scales and at galactic scales. Therefore, any progress in the analytic solution is still welcome and valuable.
    
    In this work, we propose a simplified matching scheme for the coupled first-order radial equation, derived for the Dirac field around spherically symmetric black holes. Compared to the subsequent matching across three blocks~\cite{Ternov:1980st}, our asymptotic solutions are addressed in only two blocks. An improved result in a compact and unified form is obtained with a more straightforward manipulation. Our asymptotic analysis differs from Refs.~\cite{Cotaescu:2007xw, Sporea:2015wsa} and facilitates the determination of both the real and imaginary parts of the frequency for the quasibound states. 
 In this work, we focus on the fermionic field with small mass $Mm\ll 1$. In such a case, the geodesic approximation breaks down to describe the dynamics, and the relativistic wave equation should be solved to capture the coupling between the spin and spacetime curvature.
 	
    The paper is organized as follows. We shall begin with an introduction of the master equation of fermions in Sec.~\ref{sec_equation}. Our formalism is based on gravitational quantum field theory (GQFT), and the separation of the angular and radial equations has been reconsidered within the new framework. For comparison, a brief review of the initial work~\cite{Ternov:1980st} is presented in Sec.~ \ref{sec_review}. In Sec.~\ref{sec_dirac}, we adapt the matching technique to the coupled first-order equation, which results in an \textit{improved} analytic formula for the energy spectrum. A discussion of our result and the comparison with existing results is shown in Sec.~ \ref{sec_discussion}. Finally, a concluding remark is given in Sec.~\ref{sec_summary}.
	
    \section{Master equations for the motion of fermion around spherically symmetric background based on GQFT} \label{sec_equation} 

    While the vierbein or tetrad formalism is usually adopted for treating fermions in general relativity (GR)~\cite{Cho:2003qe, Jing:2003wq, Doran:2005vm, Zhou:2013dra, Barranco:2012qs, Huang:2017nho}, we utilize the gravigauge formalism within the framework of GQFT~\cite{Wu:2015wwa,Wu:2022mzr,Wu:2022aet,Wu:2017rmd,Wu:2017urh,Wu:2024mul,Wu:2025abi} based on the internal spin gauge symmetry of the spinor field. One reason for adapting such conventions is that discussions in this paper may be regarded as a first step to solve the generalized Dirac equation there.
    
  GQFT was motivated by the goal of reconciling GR and quantum field theory (QFT) based on the fundamental premise that the laws of nature are governed by the intrinsic properties of the basic constituents of matter. This premise necessitates a rigorous distinction between intrinsic symmetries, determined by the quantum numbers of elementary particles as quantum fields, and external symmetries, which describe their motion in the flat Minkowski spacetime of coordinates. Consequently, the global Lorentz symmetry SO(1,3) in Minkowski spacetime and the intrinsic spin symmetry SP(1,3) in the Hilbert space of the Dirac spinor field are unified as joint symmetries SO(1,3)$\Join$SP(1,3) in GQFT, rather than being treated as an associated symmetry in QFT. Based on the gauge symmetry principle, the intrinsic spin symmetry SP(1,3) is localized to a gauge symmetry. To ensure the joint symmetries SO(1,3)$\Join$SP(1,3), a spin-related bicovariant vector field $\hat{\chi}_{a}^{\; \, \mu}(x)$, transforming under both spin gauge symmetry SP(1,3) and global Lorentz symmetry SO(1,3), is introduced to replace the Kronecker symbol $\delta_{a}^{\;\;\mu}$ in QFT. $\hat{\chi}_{a}^{\; \, \mu}(x)$ is regarded as an invertible vector field, and its dual bicovariant vector field $\chi_{\mu}^{\; \, a}(x)$ is shown to be a spin-related gauge-type field. 
   
   In GQFT, the fundamental gravitational field emerges as a spin-related gauge-type bicovariant vector field $\chi_{\mu}^{\; \,a}$, referred to as the gravigauge field, rather than the metric field $\chi_{\mu\nu}=\eta_{ab}\chi_{\mu}^{\; a}\chi_{\nu}^{\; b}$ in GR. The gravigauge field $\chi_{\mu}^{\; \,a}$ is regarded as a gauge-type field defined in Minkowski spacetime and valued in spin-related gravigauge spacetime, which behaves as a Goldstone-type boson, identified as a massless graviton. This conceptualizes a biframe spacetime with a fiber bundle structure, where the flat Minkowski spacetime serves as a base spacetime and the spin-related intrinsic gravigauge spacetime acts as a fiber, which is spanned by the gravigauge bases $\{ \eth_{a}\}$ and $\{\dbar \zeta^a \}$ with $\eth_a \equiv \hchi_{a}^{\; \mu} \partial_{\mu}$ and $\dbar \zeta^{a}\equiv \chi_{\mu}^{\; a}dx^{\mu}$. Especially, GQFT allows for a spin gauge invariant action term between the gravigauge field $\chi_{\mu}^{\; \, a}(x)$ and the spin gauge field $\cA_{\mu}^{ab}$, which cannot be replaced by the metric field, similar to the gravitational interactions of the Dirac spinor field. This action term introduces  nongeometric interactions that challenge the assumption of the equivalence principle in GR. Therefore, the equivalence principle in GR is no longer valid in GQFT~\cite{Gao:2024juf}. The general linear group symmetry GL(4,$\mathbb R$) that lays the foundation of GR appears as a hidden symmetry in GQFT. Unlike GR, GQFT always allows one to choose a flat Minkowski spacetime as the base spacetime. Classically, GQFT recovers GR when turning off the internal spin gauge part and neglecting the new effects due to the violation of the equivalence principle.

    Fermions are treated as the most fundamental building block in GQFT. They are governed by the generalized Dirac equation~\cite{Wu:2022aet} 
	\begin{equation}\label{eq_Dirac_origin}
		\gamma^a\hat\chi_{a}^{\; \mu}\left(i\partial_{\mu} - iV_{\mu} + \mathcal A_{\mu}\right)\Psi =\pmass\Psi
	\end{equation}
    with $\gamma^a$ being the $\gamma$-matrices and $m$ the mass of the fermion. Gravity is encoded in the gravigauge field $\chi_{\mu}^{\; a}$ and its inverse dual field $\hat{\chi}_{a}^{\; \mu}$ endowed simultaneously with coordinate index (in \textit{greek}) and spin index (in \textit{latin}). The latter one is assigned to a spin-related local orthogonal gravigauge spacetime. Hence, they are raised(lowered) by the metric $\eta^{\mu\nu}$($\eta_{\mu\nu}$) and $\eta^{ab}$($\eta_{ab}$), respectively. The spin gauge field $\mathcal A_{\mu}$ governed by the spin gauge symmetry SP(1,3) is defined as follows:
	\begin{equation}
		\mathcal A_{\mu} = \mathcal A_{\mu}^{ab}\frac12\Sigma_{ab}, \quad \Sigma_{ab}  = \frac{i}{4}[\gamma_a, \gamma_b].
	\end{equation}
    $V_{\mu}$ is an induced vector field with the following form:
    \begin{align}
    V_{\mu} &\equiv \frac{1}{2} \chi \, \hat{\chi}_{b}^{\;\; \nu}\mathcal{D}_{\nu}(\hat{\chi} \chi_{\mu}^{\;\; b}), \\
    \mathcal{D}_{\nu}(\hat{\chi}\chi_{\mu}^{\;\;b}) &= \partial_{\nu} (\hat{\chi}\chi_{\mu}^{\;\; b}) + \hat{\chi} \mathcal{A}_{\nu\, c}^{b}  \chi_{\mu}^{\;\;c}.
    \end{align}
    Here $\chi\equiv\det\chi^{\;a}_\mu$ and $\hchi = 1/\chi$. For convenience in separation of variables, the generalized Dirac equation can be recast into the following equivalent form,
    	\begin{equation}
    		\gamma^ai\hat\chi_{a}^{\; \mu}\left(\partial_{\mu} + \hat{V}_{\mu} - i\hat{A}_{\mu}\gamma^5\right)\Psi = \pmass\Psi,
    	\end{equation}
     where the vector field $\hat{V}_{\mu}$ and axial vector field $\hat{A}_{\mu}$ are defined as follows:
     \begin{align}
          \hat{V}_{\mu} &\equiv \frac{1}{2} \hat{\chi}_{b}^{\; \nu} \mathcal{A}_{\nu c}^{b}\chi_{\mu}^{\; c} - V_{\mu}, \\
          \hat{A}_{\mu} &\equiv \frac{1}{4} \epsilon_{cdc'd'} \chi_{\mu}^{\; c} \hat{\chi}^{\nu d} \mathcal{A}_{\nu}^{c'd'}. 
     \end{align}
    Detailed derivations can be found in Refs.~\cite{Wu:2015wwa, Wu:2022mzr, Wu:2022aet}. To improve the self-containment of this work and assist readers unfamiliar with GQFT, we provide a comprehensive discussion in the Appendix~\ref{app_GQFT}. 
    
    As a first step, we neglect any backreaction and quantum effects caused by spin gauge interactions and also the equivalence principle breaking effects in GQFT~\cite{Gao:2024juf}. In such an approximation, the theory recovers GR from a fundamentally different perspective. As a consequence, the background metric $g_{\mu\nu}$ of the curved spacetime is determined by the background gravigauge field $\bar{\chi}_{\mu}^{\; a}$ with the line element
    \begin{equation}\label{eq_gravi&metric}
        \mathrm ds^2 = g_{\mu\nu} dx^{\mu} dx^{\nu}, \quad g_{\mu\nu}\equiv \bar{\chi}_{\mu\nu} = \eta_{ab}\bar{\chi}_{\mu}^{\; a}\bar{\chi}_{\nu}^{\; b}.
    \end{equation} 
    In such a background, the spin gauge field is given by the spin connection
    \begin{equation}
    \begin{aligned}
        \langle \cA_{\mu}^{ab}(x) \rangle &\equiv \bOm_{\mu}^{ab}(x) \\
        &= \frac{1}{2}\left( \bhchi^{\nu a} \bfF_{\mu\nu}^{b} - \bhchi^{\nu b} \bfF_{\mu\nu}^{a} -  \bhchi^{\rho a}  \bhchi^{\sigma b}  \bfF_{\rho\sigma}^{c} \bchi_{\mu c } \right) ,
    \end{aligned}
    \end{equation}
    which is solely determined by the background gravigauge field with $\bfF_{\mu\nu}$ regarded as a field strength, 
    \begin{align}
     \bfF_{\mu\nu}^{a} =\partial_{\mu}\bchi^a_{\nu} - \partial_{\nu}\bchi^a_{\mu}.    
    \end{align}
    In this case, the corresponding vector field $\hat{V}_{\mu}$ and axial vector field $\hat{A}_{\mu}$ are calculated from the background gravigauge field as follows:
    \begin{equation}
    \begin{aligned}
        \langle \hat{V}_{\mu} \rangle &= \frac{1}{2} \bhchi_{b}^{\; \nu} \bOm_{\nu c}^{b} \bchi_{\mu}^{\; c}, \\
        \langle \hat{A}_{\mu} \rangle &= \frac{1}{4} \epsilon_{cdc'd'} \bchi_{\mu}^{\; c} \bhchi^{\nu d} \bOm_{\nu}^{c'd'} .
    \end{aligned}
    \end{equation}

    We focus on the case in which Dirac fields live outside a static spherically symmetric black hole. It is known to be the Schwarzschild black hole within the context of GR, which has an explicit background metric, 
    \begin{equation}
            \mathrm ds^2 = \frac{r-r_\text{s}}{r}\mathrm dt^2 - \frac{r\; \mathrm dr^2}{r-r_\text{s}} - r^2\left(\mathrm d\theta^2+\sin^2\theta \mathrm d\varphi^2\right) .
    \end{equation} 
    Here $r_\text{s}=2M$ denotes the Schwarzschild radius. We adapt the Planck units $G=\hbar=c=1$ throughout this article. According to Eq.~\eqref{eq_gravi&metric}, the above metric gives the following background gravigauge field:
        \begin{equation}\label{sphsym_gravigauge field}
    		\begin{aligned}
    			\bchi^0_t &= \sqrt{\frac{r-r_\text{s}}r}, &
    			\bchi^1_r &= \sqrt{\frac r{r-r_\text{s}}}, \\
    			\bchi^2_\theta &= r, &
    			\bchi^3_\varphi &= r\sin\theta ,
    		\end{aligned} 
        \end{equation}
    and all other 12 components vanish identically. The choice of the gravigauge field is fixed up to a spin gauge transformation, which originates from the intrinsic spin gauge symmetry SP(1,3) of the Dirac spinor. 
        
    In this setup, the generalized Dirac equation has the following explicit form:
    \begin{equation}\label{eq_sph_Dirac}
        \bigg[\gamma^0\sqrt{\frac{r}{r-r_\text{s}}}i\partial_t + \gamma^1\sqrt{\frac{r-r_\text{s}}{r}}iD_r + \frac{i\gamma^1\gamma^0\hat K}{r} - \pmass\bigg]\Psi=0 ,
    \end{equation}
    where we have defined, inspired by the trick work in Minkowskian QFT, the so-called $K$-operator as follows,
    \begin{equation}\label{eq_operatorK}
        \hat K = i\gamma^0\left(\gamma^1\gamma^2iD_\theta + \frac{1}{\sin\theta}\gamma^1\gamma^3i\partial_\varphi\right).
    \end{equation}
    The gamma matrices $\gamma^{0,1,2,3}$ are defined in flat spacetime and we do not introduce the gamma matrices in curved space in this work. We have introduced the following definitions:
    \begin{equation}    
     D_r \equiv \partial_r+\dfrac{3r_\text{s}-4r}{4r(r-r_\text{s})}, \quad D_\theta = \partial_\theta + \dfrac{\cot\theta}2 .
    \end{equation}
    It is easy to check that $\hat{K}$ commutes with the Dirac operator on the left-hand side of Eq.~\eqref{eq_sph_Dirac}. Furthermore, note that $[\hat K, \gamma^0]=0$, $[\hat K,\gamma^1]=0$, and also $[\hat K, \gamma^0\gamma^1] = 0$, the common eigenfunctions of $\hat K$ and $\gamma^0\gamma^1$ form a complete basis for the solutions of Eq.~\eqref{eq_sph_Dirac}. 
	
   The formalism in Eq.~\eqref{eq_operatorK} obtained in GQFT coincides with the $K$-operator in curved space~\cite{Doran:2005vm, Zhou:2013dra}, which is obtained by generalizing from flat spacetime,
    \begin{equation}
		\hat K = \frac{\gamma^0}2\left(\gamma^r\gamma^\varphi i\partial_\theta + \frac1{\sin\theta}\gamma^r\gamma^\theta i\partial_\varphi\right).
	\end{equation}
    A linear combination of gamma matrices is introduced as
   \begin{equation}
		\begin{aligned}
			\gamma^r &= \sin\theta\cos\varphi\gamma^1 + \sin\varphi\gamma^2 + \cos\theta\gamma^3, \\
			\gamma^\theta &= \cos\theta\cos\varphi\gamma^1 + \cos\varphi\gamma^2 - \sin\theta\gamma^3, \\
			\gamma^\varphi &= -\sin\varphi\gamma^1 + \cos\varphi\gamma^2 ,
		\end{aligned}
    \end{equation}
    which agrees with the definitions given in Eq.~\eqref{eq_operatorK} through a spin gauge transformation. 
	
    In Dirac representation, the $K$-operator can be written as follows,
	\begin{equation}
		\hat K = \begin{pmatrix}
			\hat{K}_+ & \\
			& \hat K_-  
		\end{pmatrix} ,
	\end{equation}
   which is block-diagonal with $\hat K_\pm$ given by 
	\begin{equation}
		\hat K_\pm = \pm\left(\sigma_3iD_\theta - \sigma_2\frac1{\sin\theta}i\partial_\varphi\right).
	\end{equation}
    The two suboperators are algebraically related by $\hat K_+ = \sigma_1\hat K_-\sigma_1$. The eigensolutions of $\hat K$ are also split into pairs of 2-spinors, with the explicit form
    \begin{align}
        \Phi^{+}_{\angparam,\azimuthal} = \begin{pmatrix} \xi_{\angparam,\azimuthal} \\ \sigma_1\xi_{\angparam,\azimuthal} \end{pmatrix}, \;
           \Phi^{-}_{\angparam,\azimuthal} = \begin{pmatrix} \xi_{\angparam,\azimuthal} \\ -\sigma_1\xi_{\angparam,\azimuthal} \end{pmatrix}.
    \end{align}
    They are also eigenstates of $\gamma^0\gamma^1$ and are intertwined as shown by the following algebraic properties
    \begin{equation}\label{eq_intertwiners}
        \begin{aligned}
            \gamma^0\Phi^{\pm}_{\angparam,\azimuthal} &=  \Phi^{\mp}_{\angparam,\azimuthal}, \\
            \gamma^1\Phi^{\pm}_{\angparam,\azimuthal} &= \pm\Phi^{\mp}_{\angparam,\azimuthal}, \\
            \gamma^0\gamma^1\Phi^{\pm}_{\angparam,\azimuthal} &=  \pm\Phi^{\pm}_{\angparam,\azimuthal}.
        \end{aligned}
    \end{equation}
    The 2-spinor $\xi_{\angparam,\azimuthal}$ is the eigenstate of the operator $\hat K_+$,
    \begin{equation}\label{eq_angular_component}
        \hat K_+\xi_{\angparam,\azimuthal}(\theta,\varphi)=\angparam\xi_{\angparam,\azimuthal}(\theta,\varphi) .
    \end{equation}
    Here $\angparam$ are eigenvalues of the operator $\hat K_+$ and $\azimuthal$ the azimuthal quanta. 
    
   The spinor wave functions have a $4\pi$-periodicity in $\varphi$ and satisfy the following boundary conditions 
    \begin{gather}
		\xi_{\angparam,\azimuthal}(\theta,\varphi+4\pi) = \xi_{\angparam,\azimuthal}(\theta,\varphi), \\
		\partial_\varphi\xi_{\angparam,\azimuthal}|_{\theta=0} = \partial_\varphi\xi_{\angparam,\azimuthal}|_{\theta=\pi} = 0.
    \end{gather}
    Therefore solutions of Eq.~\eqref{eq_angular_component} can be expressed as 
    \begin{equation}
        \begin{aligned}
            \xi_{\angparam,\azimuthal}(\theta,\varphi) \propto e^{i\azimuthal\varphi}\bigg[\frac{ F^{\angparam,-\angparam}_{\frac12-\azimuthal}(\sin^2\frac\theta2)}{\angparam\tan^{\azimuthal}(\frac\theta2)\sqrt{\sin\theta}}\begin{pmatrix}1\\-1\end{pmatrix} & \\
            - \dfrac{\sin\theta F^{1+\angparam,1-\angparam}_{\frac32-\azimuthal}(\sin^2\frac\theta2)}{1-2\azimuthal}\begin{pmatrix}1\\1\end{pmatrix}\bigg] &.
        \end{aligned}
    \end{equation}
    Here $F^{\alpha,\beta}_\gamma(z)$ denotes the hypergeometric function of variable $z$. The eigenvalues $\angparam$ and $\azimuthal$ can be determined with the boundary conditions
    \begin{align}
        \angparam = \pm \absangparam, & &
        \azimuthal = \pm\frac12, \pm\frac32, \dots, \pm\absangparam-\frac12,
    \end{align}
    with $\absangparam$ an arbitrary positive integer. 

    With the angular wave functions $\Phi^\pm_{\angparam,\azimuthal}$, the Dirac field in Eq.~\eqref{eq_sph_Dirac} can be expressed as 
    \begin{equation}
        \begin{aligned}
            \Psi(t,r,\theta,\varphi) = \frac{\mathrm e^{-i\omega t}}{r^\frac{3}{4}(r-r_\text{s})^\frac{1}{4}} \bigg(\psi^+_{\omega,\angparam}(r)\Phi^+_{\angparam,\azimuthal}(\theta,\varphi) &\\
            + \psi^-_{\omega,\angparam}(r)\Phi^-_{\angparam,\azimuthal}(\theta,\varphi)\bigg) &,
        \end{aligned}
    \end{equation}
    where $\psi^\pm_{\omega,\angparam}$ are radial wave functions and $\omega$ denotes the eigenfrequency relevant to the energy of the field. For a compact expression, we also adapt the notation 
    \begin{equation}
        \psi_{\omega,\angparam}(r)=\begin{pmatrix}
            \psi^+_{\omega,\angparam}(r) \\ \psi^-_{\omega,\angparam}(r)
        \end{pmatrix}
    \end{equation}
    in the following discussion. Substituting into Eq.~\eqref{eq_sph_Dirac} and using the properties of $\Phi^{\pm}_{\angparam,\azimuthal}$ in Eq.~\eqref{eq_intertwiners}, we arrive at the following coupled radial equations:
    \begin{equation}\label{eq_coupled_equation}
		\begin{aligned}
			\left(\partial_{r} - \dfrac{i\omega r}{r-1}\right)\psi^{+}_{\omega,\angparam}(r) &= 
			-\frac{\angparam+i\pmass{r}}{\sqrt{{r}({r}-1)}}\psi^{-}_{\omega, \angparam}(r), \\
			\left(\partial_{r} + \dfrac{i\omega r}{r-1}\right)\psi^{-}_{\omega, \angparam}(r) &= -\frac{\angparam-i\pmass{r}}{\sqrt{{r}({r}-1)}}\psi^{+}_{\omega, \angparam}(r).
		\end{aligned}
    \end{equation}
   Here and in the following discussion we take $r_\text{s}=1$ without any loss of generality. The eigenfrequency $\omega$ can be determined by solving these equations with quasibound state boundary conditions that the field is purely ingoing at the horizon and decays exponentially at infinity.
    
   The radial equations~\eqref{eq_coupled_equation} are consistent with the equations presented in~\cite{Unruh:1976fm}, derived from the vierbein formalism. From now on, we should focus on such a radial equation for given $\omega$, $\azimuthal$, $\angparam$. When no confusion arises, subscripts are ignored in the following discussion for clarity.
	
    \section{Matching technique in the second-order equation of motion}
    \label{sec_review}

    As Eq.~\eqref{eq_coupled_equation} is coupled, a straightforward approach is to solve the second-order equation for one of the components. Let us temporarily focus on the case of $m_\ell=\ell$, where we arrive at the following equation:
    \begin{equation}\label{eq_quad}
		\bigg(\partial_r^2 + P_\pm(r)\partial_r + Q_\pm(r)\bigg)\psi^\pm = 0,
    \end{equation}
    where the coefficients are defined as follows:
	\begin{equation} 
		\begin{aligned}
		  P_+(r) &= \frac1{2r} + \frac1{2(r-1)} - \frac1{r-i\absangparam/\pmass}, \\
		  Q_+(r) &= -p^2 + \frac{\absangparam^2}{r} - \frac{\absangparam^2+\pmass^2-2\omega^2+\frac{\absangparam\omega/\pmass}{1-i\absangparam/\pmass}}{r-1} \\ 
          &\qquad + \frac{\omega^2+i\omega/2}{(r-1)^2} + \frac{\absangparam\omega/\pmass}{1-i\absangparam/\pmass}\frac{1}{r-i\absangparam/\pmass}, \\
		\end{aligned}
	\end{equation}
    with $p=\sqrt{\pmass^2-\omega^2}$. Since $P_- =P_+^*$, $Q_-=Q_+^*$, the two components are related via $\psi^-\sim\psi^{+*}$. Therefore they yield the same spectrum of eigenfrequencies. In the following parts, we will focus on the component $\psi^+$.

 The differential equation~\eqref{eq_quad} possesses three regular singularities $r=0$, $r=1$, $r=i\ell/{m}$ and one irregular singularity $r=+\infty$. In the scenario with $\omega\sim\pmass\ll1$, approximate solutions in neighborhoods of $r= 1$, $r= \absangparam/\pmass$ and $r\gg \absangparam/\pmass$ can be derived, which are denoted as $\psi_h$, $\psi_i$ and $\psi_f$, respectively. These solutions are subsequently matched in the overlapping regions to determine the eigenfrequency $\omega$, as an expansion in a series of $\pmass$. Such a technique is referred to as block-matching or matched asymptotic expansion. It was first applied in~\cite{Ternov:1980st} and the result is consistent with the numerical calculation~\cite{Dolan:2015eua}.
    
    In the \textit{far} region $r>{\absangparam}/{\pmass}$, Eq.~\eqref{eq_quad} is approximately
    \begin{equation}\label{eq_quad_approx_far}
        \bigg(\partial_r^2-p^2 + \frac{2\omega^2-\pmass^2}{r} - \frac{\tilde{\absangparam}(\tilde{\absangparam}+1)}{r^2}\bigg)\psi^+_f = 0 ,
    \end{equation}
    with $\tilde{\absangparam}=\absangparam+\epsilon_\absangparam$. Here $\epsilon_\absangparam$ is small but cannot be dropped directly \cite{Bao:2022hew, Bao:2023xna,Chu:2024iie}. It serves as a regulator for the  ill-defined Gamma functions in the result as shown in Eq.~\eqref{eq_delta_quad} and results in a factor of 1/2 missed in the original result~\cite{Ternov:1980st}. Such an equation with a vanishing boundary condition at infinity can be solved by the Whittaker function,
    \begin{equation}\label{eq_psif}
        \psi^+_f \propto W_{\tomega,\tilde{\absangparam}+\frac12}(2pr) , \quad \tomega \equiv \frac{2\omega^2-\pmass^2}{2p},
    \end{equation}
    for $p>0$.
    
     To figure out the approximate solution in the region \textit{near} the horizon $1\leq r<\absangparam/{m}$, we decompose $\psi^+_h(r)$ as
     \begin{equation}
         \psi^+_h(r) = r^{i\omega+1/2}(r-1)^{-i\omega+1/2}\hat\psi^+_h(r).
     \end{equation}
    Then the equation for $\hat\psi^+_h(r)$ is given by
    \begin{equation} 
		\bigg[r(r-1)\partial_r^2 + \left(3r - \frac32 - 2i\omega\right)\partial_r - \tilde{\absangparam}^2 + 1\bigg]\hat\psi^+_h = 0 .
    \end{equation}
   When accounting for the purely ingoing boundary condition at the horizon, the solution is given in terms of a hypergeometric function
    \begin{equation}\label{eq_psih}
		\psi^+_h( r) \propto \left(\frac{r}{r-1}\right)^{i\omega}\sqrt{r(r-1)}F^{1-\tilde{\absangparam}, 1+\tilde{\absangparam}}_{\frac32-2i\omega}(1- r).
    \end{equation}
  By utilizing $1-r$ as the variable of the hypergeometric function instead of $r$ as in Ref.~\cite{Ternov:1980st}, we can easily see that the solution in Eq.~\eqref{eq_psih} is a purely incoming wave at the horizon.
	
\begin{figure}[t]
        \centering
        \includegraphics[width=0.5\textwidth]{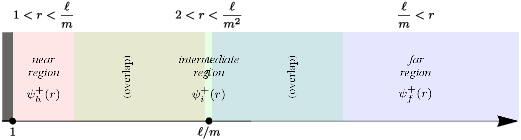} \\
        \includegraphics[width=0.5\textwidth]{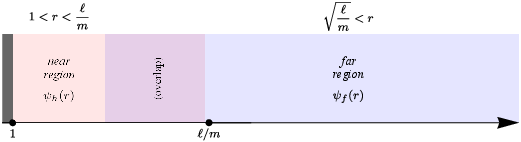}
        \caption{Different regions used for the asymptotic solutions. Upper: the three regions for the sequential matching scheme. The {\it near} and {\it far} regions are separated by $r=\ell/m$. Bottom: the two regions for the direct matching scheme for the coupled first-order equation.}
        \label{fig_illustration}
    \end{figure}

    To determine the eigenfrequency, we need to match $\psi^+_h$ and $\psi^+_f$ to construct a solution that simultaneously satisfies the boundary conditions at the horizon ($r=1$) and at spatial infinity ($r=\infty$). However, the two solutions are only valid in the nonoverlapping regions $r<\ell/\pmass$ and $r>{\absangparam}/{\pmass}$, respectively. To accommodate this requirement, it is useful to introduce an intermediate region $2<r\sim\ell/m< {\ell}/{m^2}$ as a bridge, as illustrated in Fig.~\ref{fig_illustration}. Within this region, the wave function $\psi_i^+$ is determined via the following equation:
    \begin{equation} 
        \bigg[r\left(r-\frac{i\tilde{\absangparam}}\pmass\right)\partial_r^2 - \frac{i\tilde{\absangparam}}{\pmass}\partial_r - \tilde{\absangparam}(\tilde{\absangparam}+1) + \frac{i\tilde{\absangparam}^3}{\pmass r}\bigg]\psi^+_i(r) = 0.
    \end{equation}
    Its solution is given by
    \begin{equation}\label{eq_psii}
        \begin{aligned}
            \psi_i^+(r) &= c_1\left(\frac{\pmass r}{i\tilde{\absangparam}}\right)^{\tilde{\absangparam}}\left(1+2\tilde{\absangparam}+2i\pmass r\right) + c_2\left(\frac{\pmass r}{i\tilde{\absangparam}}\right)^{-\tilde{\absangparam}} ,
        \end{aligned}
    \end{equation}
    where the coefficients $c_1$ and $c_2$ are two constants of integration. The term proportional to $c_1$ behaves as $r^\ell$ when approaching the horizon, while it behaves as $r^{\ell+1}$ when approaching spatial infinity. The distinct asymptotic behaviors are crucial for the matching with the solutions in the near and far regions.

    Although the near and far regions do not overlap with each other, they each overlap with the intermediate region separately. Within the overlapping of the near and intermediate regions, the local solutions $\psi_h$ and $\psi_i$ get the following asymptotic forms:
    \begin{equation} 
		\begin{aligned}
            \psi^+_h(r) &\rightarrow r^{\tilde{\absangparam}}\frac{\Gamma(2\tilde{\absangparam})\Gamma(\frac32-2i\omega)}{\Gamma(\frac12+\tilde{\absangparam}-2i\omega)\Gamma(1+\tilde{\absangparam})} \\
            &\qquad\ + r^{-\tilde{\absangparam}}\frac{\Gamma(-2\tilde{\absangparam})\Gamma(\frac32-2i\omega)}{\Gamma(\frac12-\tilde{\absangparam}-2i\omega)\Gamma(1-\tilde{\absangparam})}
		\end{aligned}
    \end{equation}
    for $r\gg1$ and
    \begin{equation}
        \begin{aligned}
            \psi^+_i(r) &\rightarrow c_1\left(\frac{\pmass r}{i\tilde{\absangparam}}\right)^{\tilde{\absangparam}}(1+2\tilde{\absangparam}) + c_2\left(\frac{\pmass r}{i\tilde{\absangparam}}\right)^{-\tilde{\absangparam}}
        \end{aligned}
    \end{equation}
    for $r\ll{\absangparam}/{m}$. These two local solutions should exhibit the same behavior and result in the following matching condition,
    \begin{equation}\label{eq_cond1}
		\begin{aligned}
            \frac{(\pmass/i\tilde{\absangparam})^{-2\tilde{\absangparam}}}{2\tilde{\absangparam}+1}\frac{c_2}{c_1} &= -\frac{\Gamma(-2\tilde{\absangparam})}{\Gamma(-\tilde{\absangparam})}\frac{\Gamma(\tilde{\absangparam})}{\Gamma(2\tilde{\absangparam})}\frac{\Gamma(\frac12+\tilde{\absangparam}-2i\omega)}{\Gamma(\frac12-\tilde{\absangparam}-2i\omega)} .
		\end{aligned}
    \end{equation}
    
    On the other hand, within the interval $\absangparam/\pmass\ll r\ll\absangparam/{m^2}$, the solutions $\psi_f$ and $\psi_i$ are asymptotically given by
    \begin{equation}\label{eq_f2i}
		\begin{aligned}
            \psi^+_f(r) \rightarrow &\ (2pr)^{\tilde{\absangparam}+1}\frac{\Gamma(-1-2\tilde{\absangparam})}{\Gamma(-\tilde{\absangparam}-\tomega)} \\
            &\ + (2pr)^{-\tilde{\absangparam}}\frac{\Gamma(1+2\tilde{\absangparam})}{\Gamma(1+\tilde{\absangparam}-\tomega)} \\
		\end{aligned}
    \end{equation}
    for $r\ll\absangparam/{m^2}$ and 
    \begin{equation}
        \begin{aligned}
            \psi^+_i(r) &\rightarrow c_1\left(\frac{\pmass r}{i\tilde{\absangparam}}\right)^{\tilde{\absangparam}+1}2\tilde{\absangparam} + c_2\left(\frac{\pmass r}{i\tilde{\absangparam}}\right)^{-\tilde{\absangparam}} \\
        \end{aligned}
    \end{equation}
    for $r\gg\absangparam/\pmass$. The matching of two functions results in the following condition:
	\begin{equation}
        \begin{aligned}
            -\frac{2i\pmass}{(2p)^{2\tilde{\absangparam}+1}}\frac{\Gamma(1+2\tilde{\absangparam})\Gamma(-\tilde{\absangparam}-\tomega)}{\Gamma(-2\tilde{\absangparam})\Gamma(1+\tilde{\absangparam}-\tomega)} = \frac{(\pmass/i\absangparam)^{-2\tilde{\absangparam}}}{2\tilde{\absangparam}+1}\frac{c_2}{c_1}.
        \end{aligned}
	\end{equation}
    Together with Eq.~\eqref{eq_cond1}, we arrive at the following equation for $\omega$:
    \begin{equation}\label{eq_cond_complete}
        \begin{aligned}
            \frac{2i\pmass}{(2p)^{2\tilde{\absangparam}+1}}&\frac{\Gamma(2\tilde{\absangparam}+1)\Gamma(-\tilde{\absangparam}-\tomega)}{\Gamma(-2\tilde{\absangparam})\Gamma(\tilde{\absangparam}-\tomega +1)}  \\
            &\ = \frac{\Gamma(-2\tilde{\absangparam})}{\Gamma(-\tilde{\absangparam})}\frac{\Gamma(\tilde{\absangparam})}{\Gamma(2\tilde{\absangparam})}\frac{\Gamma(\frac12+\tilde{\absangparam}-2i\omega)}{\Gamma(\frac12-\tilde{\absangparam}-2i\omega)} ,
        \end{aligned}
    \end{equation}
    which enables us to determine the spectrum of fermions in a spherically symmetric black hole background. 
	
    The above equation can be solved perturbatively. 
     Since the field is confined in the potential well outside the horizon,  as proposed in Ref.~\cite{Detweiler:1980uk}, we expect Eq.~\eqref{eq_f2i} to remain convergent in the small $r$ region and  the coefficient of $r^{-\tilde{\absangparam}}$ to be suppressed, which requires the following condition:
    \begin{equation}\label{eq_quantized}
		\tilde{\absangparam}-\tomega +1 = -n+\delta, \quad n=0,1,2,\dots,  
    \end{equation}
    where $\delta$ is complex and small. The eigenfrequency admits an expansion $\omega \approx \omega_0 + \omega_1\delta$. Given that $\tomega$ is a function of $\omega$, both $\omega_0$ and $\omega_1$ can be solved from Eq.~\eqref{eq_quantized} as follows,
    \begin{equation}
		\begin{aligned}
            \omega_0 &\approx \pmass\left[1 - \dfrac{\pmass^2}{8\bar n^2} + \mathcal O(\pmass^4)\right], \\
            \omega_1 &= -\frac1{\omega_0}\left(\frac{3\pmass^2-2\omega_0^2}{2p_0^{3}} - \frac{\absangparam}{\pmass\omega_0\tilde{\absangparam}}\right)^{-1} \\
            &\approx -\frac{\pmass^3}{4\bar n^3} + \mathcal O(\pmass^4) ,
		\end{aligned}
    \end{equation}
    where $\bar n=n+\absangparam+1$ and $p_0=\sqrt{\pmass^2-\omega_0^2}$. Substituting Eq.~\eqref{eq_quantized} into Eq.~\eqref{eq_cond_complete}, we arrive at a concrete expression for the small quantity $\delta$,
    \begin{equation}\label{eq_delta_quad}
		\begin{aligned}
            \delta \approx \frac{(2p_0)^{2\tilde{\absangparam}+1}}{2i\pmass}\frac{\Gamma(\tilde{\absangparam})}{n!\Gamma^2(2\tilde{\absangparam}+1)}\frac{\Gamma(\frac12+\tilde{\absangparam}-2i\omega_0)}{\Gamma(\frac12-\tilde{\absangparam}-2i\omega_0)}& \\
             \times\frac{(-)^{n}\Gamma(1-2\tilde{\absangparam})}{\Gamma(-n-2\tilde{\absangparam}-1)}\frac{\Gamma(-2\tilde{\absangparam})}{\Gamma(-\tilde{\absangparam})} ,
		\end{aligned}
    \end{equation}
    which leads to an extremely small value for $\delta$, i.e., $\delta = \mathcal O(\pmass^{4\tilde{\absangparam}+1})$. 
    
    In the limit of $\epsilon_\absangparam\to 0$, some Gamma functions in the second line of Eq.~\eqref{eq_delta_quad} are ill-defined, while the whole expression remains meaningful with the following observation:
    \begin{equation}\label{eq_factor2}
		\begin{aligned}
            \lim_{\epsilon_\ell \rightarrow 0}\frac{\Gamma(-2\tilde{\absangparam})}{\Gamma(-\tilde{\absangparam})} &= \frac{(-)^{\absangparam}}2\frac{\absangparam!}{(2\absangparam)!}, \\
            \lim_{\epsilon_\ell \rightarrow 0}\frac{\Gamma(1-2\tilde{\absangparam})}{\Gamma(-n-2\tilde{\absangparam}-1)} &= (-)^{n}\frac{(n+2\absangparam+1)!}{(2\absangparam-1)!}.
		\end{aligned}
	\end{equation}
Here we can see that $\epsilon_\ell$ helps to regularize the Gamma functions and we can recover the result in Refs.~\cite{Ternov:1980st, Dolan:2015eua} by taking $\epsilon_\ell\to 0$.  We will show in the next section that the nonzero $\epsilon_\absangparam$ not only acts as a regulator but also provides sizable correction to the spectrum. 

    Combining the above ingredients, we arrive at the following relation for the imaginary part of the spectrum,
    \begin{equation}\label{eq_imagspec_quad_positive}
        \begin{aligned}
            \frac{\text{Im }\omega_{\angparam>0}}{\pmass}\approx -\frac{(\pmass M)^{4\absangparam+3}}{(n+\absangparam+1)^{2\absangparam+4}}\frac{(n+2\absangparam+1)!}{n![(2\absangparam)!]^2} & \\
            \times\prod_{j=0}^{\absangparam-1}\left[1+\left(\frac{4\omega_0 M}{j+1/2}\right)^2\right] &, 
        \end{aligned}
    \end{equation}
    where the unit $r_\text{s}=2M$ has been recovered. 
 
    For the case $\angparam=-\absangparam$, the spectrum can be obtained with the same procedure as in the case $\angparam=\absangparam$. The result is slightly different,
    \begin{equation}\label{eq_imagspec_quad_negative}
        \begin{aligned}
            \frac{\text{Im }\omega_{\angparam<0}}{\pmass}\approx -\frac{(\pmass M)^{4\absangparam+1}}{(n+\absangparam)^{2\absangparam+2}}\frac{(n+2\absangparam-1)!}{n![(2\absangparam-1)!]^2} & \\
            \times\prod_{j=0}^{\absangparam-1}\left[1+\left(\frac{4\omega_0 M}{j+1/2}\right)^2\right] &.
        \end{aligned}
    \end{equation}
    The results Eqs.~\eqref{eq_imagspec_quad_positive} and~\eqref{eq_imagspec_quad_negative} coincide with those obtained in Ref.~\cite{Dolan:2015eua} by replacing $n \rightarrow n + \ell + \dfrac{s+1}{2}$ and $\ell \rightarrow \ell - \dfrac{1-s}2$ in the literature. It is noteworthy that a common misconception regarding the ill-defined Gamma functions, as identified in~\cite{Bao:2022hew}, was present in the early work~\cite{Ternov:1980st}. The missing factor of $1/2$ has been rectified in Ref.~\cite{Dolan:2015eua}.
    
    \section{Simplified matching and improved analytic spectrum }
    \label{sec_dirac}
	
	In the previous section, we outlined the matching technique applied to the second-order equation of motion derived from the Dirac equation in the background of a Schwarzschild black hole. It was found that solutions near the horizon and in the far region cannot be matched directly.  To surmount this impediment, we now shift our attention to the coupled radial equation~\eqref{eq_coupled_equation} to which the block-matching technique can be applied directly.

	Unlike the second-order equation~\eqref{eq_quad}, the equations in Eq.~\eqref{eq_coupled_equation} admit only three singularities $r=0$, $r=1$ (horizon) and $r=+\infty$. In the \textit{far} region $r\gg 1$, retaining terms up to order $\mathcal O(r^{-1})$,  Eq.~\eqref{eq_coupled_equation} reduces to the following asymptotic form,
    \begin{equation}\label{eq_far}
		\begin{aligned}
			\left(\partial_{r} - i\omega - \frac{i\omega}{r}
			\right)\psi^+_f + \left(i\pmass + \frac{\angparam+i\pmass/2}{
				r}\right)\psi^-_f &= 0, \\
			\left(\partial_{r} + i\omega + \frac{i\omega}{r}
			\right)\psi^-_f - \left(i\pmass - \frac{\angparam-i\pmass/2}{
				r}\right)\psi^+_f &= 0.
		\end{aligned}
    \end{equation}
    Our primary objective is to match the above solution in the far region with the solution near the horizon. To achieve our goal, it is crucial to reveal the behavior of the function $\psi_f$ in the small-$r$ limit. 
	
	To diagonalize the coefficient matrix associated with $1/r$, it is useful to introduce the following transformation,
    \begin{equation} 
		\begin{aligned}
            \tilde\psi^+_f &= \left(\frac12{+}\frac{i\omega}{2\tilde{\absangparam}}\right)\psi^{{-}}_f + \frac{\angparam-i\pmass/2}{2\tilde{\absangparam}}\psi^{+}_f, \\
            \tilde\psi^-_f &= \left(\frac12{-}\frac{i\omega}{2\tilde{\absangparam}}\right)\psi^{{-}}_f - \frac{\angparam-i\pmass/2}{2\tilde{\absangparam}}\psi^{{+}}_f,
		\end{aligned}
    \end{equation} 
    which allows us to obtain the following simplified equations:
    \begin{equation}
		\begin{aligned}
		  \left(\partial_{r} + \frac{\tilde{\absangparam}}{r} - \frac{p\tomega}{\tilde{\absangparam}}\right)\tilde\psi^+_f(  r) &= -i\beta_+\tilde\psi^-_f(  r), \\
		  \left(\partial_{r} - \frac{\tilde{\absangparam}}{r} + \frac{p\tomega}{\tilde{\absangparam}}\right)\tilde\psi^-_f(  r) &= -i\beta_-\tilde\psi^{+}_f(r),
		\end{aligned}
    \end{equation}
    with the notations:
    \begin{align}
        \tilde{\absangparam} & =\sqrt{\angparam^2+\frac{\pmass^2}{4}-\omega^2} \equiv \ell+\epsilon_\absangparam, \label{eq_effective_angparam} \\
        \beta_\pm &= \omega\dfrac{2\angparam+i\pmass}{2\angparam-i\pmass}\left(1\pm\dfrac{i\omega}{\tilde{\absangparam}}\right) \pm \pmass\dfrac{\angparam}{\tilde{\absangparam}}.\label{eq_beta_pm}
    \end{align}
    The approximate solution in the far region is presented in terms of a Whittaker function as follows,
    \begin{equation}\label{eq_psi_h_Dirac}
		\tilde\psi_f(r) \propto 
		\begin{pmatrix}
		  \frac{p(\tilde{\absangparam}-\tomega)}{i\tilde{\absangparam}\beta_-}W_{\tomega,\tilde{\absangparam}+\frac12}(2pr) \\
		  W_{\tomega,\tilde{\absangparam}-\frac12}(2pr) 
		\end{pmatrix},
    \end{equation}
    where we have imposed the boundary condition $\tilde\psi_f\rightarrow0$ as $r\rightarrow\infty$. 
	
	The eigenfrequencies are determined by the boundary conditions.
	The solution $\psi_f$ (or equivalently $\tilde\psi_f$) obtained in the far region cannot capture the boundary condition at $r=1$. Therefore, we need one more solution near the horizon to complete the matching. In the \textit{near} region, with the assumption $\tilde{\ell}\simeq \ell$ and $\omega\sim m\ll 1$,   the coupled equations in Eq.~\eqref{eq_coupled_equation} can be rewritten as
    \begin{equation}\label{eq_horizon}
		\begin{aligned}
            \left(\sqrt{r(r-1)}\partial_r - \frac{i\omega}{\sqrt{r(r-1)}}\right)\psi^+_h + s\tilde{\absangparam}\psi^-_h &= 0, \\
            \bigg(\sqrt{r(r-1)}\partial_r + \frac{i\omega}{\sqrt{r(r-1)}}\bigg)\psi^-_h + s\tilde{\absangparam}\psi^+_h &= 0 ,
		\end{aligned}
    \end{equation}
    with $s=\text{sgn }(\angparam)$. In the allowed range, it is the energy rather than the mass that is crucial for dynamics. The approximate solutions are presented in terms of hypergeometric functions, as follows,
    \begin{equation}\label{eq_sols_horizon}
		\psi_h(r) =\left(\frac{r}{r-1}\right)^{i\omega}
		\begin{pmatrix}
			\sqrt{r(r-1)}F^{1+\tilde{\absangparam},1-\tilde{\absangparam}}_{\frac{3}{2}-2i\omega}(1-r) \\ 
			\frac{s}{\tilde{\absangparam}}\left(2i\omega-\frac12\right)F^{\tilde{\absangparam},-\tilde{\absangparam}}_{\frac{1}{2}-2i \omega}(1-r)
		\end{pmatrix},
    \end{equation}
    which is consistent with the solution in Eq.~\eqref{eq_psih} and accounts for the boundary condition that the solution is purely ingoing at the horizon $r=1$. 
    
    Equation~\eqref{eq_far} is valid for $r>\sqrt{\ell/m}$ while Eq.~\eqref{eq_horizon} holds for $r<\ell/m$ in the case of $\ell/m\gg1$, and there exists an overlapping region where both equations are valid. The nonemptiness of this overlap emphasizes the light-particle assumption $\omega\sim \pmass\ll1$. In the overlapping region, the approximate solution in the near region can be described by its asymptotic expansion with $r\gg 1$,
        \begin{equation} 
		\begin{aligned}
            \psi_h(r) \approx r^{-\tilde{\absangparam}}\frac{\Gamma(-2\tilde{\absangparam})\Gamma(\frac32-2i\omega)}{\Gamma(1-\tilde{\absangparam})\Gamma(\frac12-\tilde{\absangparam}-2i\omega)}\begin{pmatrix}1 \\ s\end{pmatrix} & \\
            + r^{\tilde{\absangparam}}\frac{\Gamma(2\tilde{\absangparam})\Gamma({\frac32}-2i\omega)}{\Gamma(1+\tilde{\absangparam})\Gamma(\frac12+\tilde{\absangparam}-2i\omega)}\begin{pmatrix}1 \\ -s\end{pmatrix} &.
		\end{aligned} 
	\end{equation}
    And for $r\ll\absangparam/{\pmass}$, the solution $\tilde\psi_f$ can be described by
    \begin{equation}\label{eq_psif2h}
        \begin{aligned}
    		\tilde\psi_f(r) \approx (2pr)^{-\tilde{\absangparam}}\frac{2p}{i\beta_-}\frac{\Gamma(2\tilde{\absangparam})}{\Gamma(\tilde{\absangparam}-\tomega)}\begin{pmatrix}1\\0\end{pmatrix} & \\
    		+ (2pr)^{\tilde{\absangparam}}\frac{\Gamma(1-2\tilde{\absangparam})}{\Gamma(1-\tilde{\absangparam}-\tomega)}
    		\begin{pmatrix}0\\1\end{pmatrix} &.
        \end{aligned}
    \end{equation}
    Next, we consider the case with $\angparam=-\absangparam$. The coefficient of $r^{-\tilde{\absangparam}}$ should be highly suppressed, which requires the following condition,
    \begin{equation}\label{eq_quan_cond_Dirac}
		\tilde{\absangparam}-\tomega = -n+\delta, \quad n=0,1,2,\dots,
	\end{equation}
 and $\delta$ is a small and complex quantity in which the eigenfrequency can be expressed as $\omega\approx\omega_0+\omega_1\delta$. 
    
    The above equation leads to the following solutions:
    \begin{equation}\label{eq_Dirac_auxilliary}
		\begin{aligned}
            \omega_0 &\approx \pmass\left[1 - \frac{\pmass^2}{8\bar n^2} + \left[\frac{15}{8}-\frac{3\bar n}{2\absangparam}\right]\frac{\pmass^4}{16\bar n^4} \right],\\ 
            \omega_1 &= -\frac{1}{\omega_0}\left(\frac{3\pmass^2-2\omega_0^2}{2p_0^3}+\frac1{\tilde{\absangparam}}\right)^{-1} \approx -\frac{\pmass^3}{4\bar n^3}, \\
            p_0 &= \sqrt{\pmass^2-\omega_0^2}\approx\dfrac{\pmass^2}{2\bar n},
		\end{aligned}
    \end{equation}
    with $\bar n=n+\absangparam$. To proceed, it is necessary to switch from $\tilde\psi$ to $\psi$. Expanding around $\omega_0$ and keeping the leading term of $\delta$, we arrive at the following expression:
    \begin{equation}
		\begin{aligned}
            \psi_f(r) \approx\ (2pr)^{-\tilde{\absangparam}}\frac{2p}{i\beta_-}(-)^{n}n!\Gamma(2\tilde{\absangparam})\delta\begin{pmatrix}1\\s\end{pmatrix} & \\
            + (2pr)^{\tilde{\absangparam}}\frac{\Gamma(1-2\tilde{\absangparam})}{\Gamma(-n-2\tilde{\absangparam}+1)}\begin{pmatrix}1\\-s\end{pmatrix}&.
        \end{aligned}
    \end{equation}
    By matching the two local solutions, we obtain the result for $\delta$ as follows:
    \begin{align}\label{eq_delta_Dirac}
		\delta \approx  i\beta_-(2p_0)^{2\tilde{\absangparam}-1}\frac{\Gamma(\tilde{\absangparam})}{n!\Gamma^2(2\tilde{\absangparam})}
		\frac{\Gamma(\frac12+\tilde{\absangparam}-2i\omega_0)}{\Gamma(\frac12-\tilde{\absangparam}-2i\omega_0)} &\nonumber \\
         \times
		\frac{\Gamma(-2\tilde{\absangparam})}{\Gamma(-\tilde{\absangparam})}
		\frac{(-)^{n+1}\Gamma(1-2\tilde{\absangparam})}{\Gamma(-n-2\tilde{\absangparam}+1)}& ,
    \end{align}
    which recovers Eq.~\eqref{eq_imagspec_quad_negative} in the limit $\tilde{\absangparam}\rightarrow \absangparam$. 
    
    The above result is valid for $m_\ell=\pm \ell$, except the case of $n=0$. For $m_\ell =\ell$ and $n=0$, we can obtain $\beta_-\sim \mathcal{O}(m^3\delta)$ from Eq.~\eqref{eq_beta_pm} with Eq.~\eqref{eq_quan_cond_Dirac}, and the coefficient of $r^{-\tilde\absangparam}$ in Eq.~\eqref{eq_psif2h},
    \begin{equation}
        \frac{2p\Gamma(2\tilde{\absangparam})}{\beta_-\Gamma(\tilde{\absangparam}-\tomega)} \approx \mathcal O\left(\frac{1}{\pmass}\right),
    \end{equation}
    which contradicts our earlier assumption regarding the coefficient of $r^{-\tilde{\absangparam}}$ in Eq.~\eqref{eq_psif2h}. Thus, for a positive constant $\angparam=\absangparam$, the quantum number $n$ should start from 1 rather than 0.  To incorporate the two different cases, it is preferable to adopt the following replacement:
    \begin{equation}\label{eq_n_replacement}
        n\rightarrow n+\frac{1+s}2 ,
    \end{equation}
    which accounts for the sign of $\absangparam$ in all the results obtained from the above Eqs.~\eqref{eq_Dirac_auxilliary} and~\eqref{eq_delta_Dirac}. Thereby, we can successfully retrieve the result in Eq.~\eqref{eq_imagspec_quad_positive} in the limit $\tilde{\absangparam}\rightarrow \absangparam$.
	
    Therefore, it enables us to obtain the final result in a unified form as follows:
    \begin{equation}\label{eq_improved_Dirac}
		\begin{aligned}
            \frac{\omega}{\pmass} &\approx 1 - \frac{(\pmass M)^2}{2\bar n^2} + \left(\frac{15}{8}-\frac{3\bar n}{2\absangparam_0}\right)\frac{(\pmass M)^4}{\bar n^4} \\ 
            &\ + i\beta_-\left(\frac{3\pmass^2-2\omega_0^2}{2p_0M}+\frac{2p_0^2}{\tilde{\absangparam_0}}\right)^{-1}\frac{(4p_0M)^{2\tilde{\absangparam_0}+1}}{16M^3\pmass\omega_0}\Gamma_{n\tilde{\absangparam_0} s},
		\end{aligned}
    \end{equation}
    in terms of the quantum number
    \begin{equation}
      \bar{n} = n + \absangparam + \dfrac{s+1}2, \qquad n=0,1,2,\dots,
    \end{equation}
    and $\Gamma_{n\tilde{\absangparam_0} s}$ is explicitly given by
    \begin{widetext}\begin{equation}
        \Gamma_{n\tilde{\absangparam_0} s} = \frac{(-)^{n+\frac{1-s}2}}{(n+\frac{1+s}2)!}\frac{\Gamma(\tilde{\absangparam_0})}{\Gamma^2(2\tilde{\absangparam_0})}\frac{\Gamma(\frac12+\tilde{\absangparam_0}-4i\omega_0M)}{\Gamma(\frac12-\tilde{\absangparam_0}-4i\omega_0M)}\frac{\Gamma(1-2\tilde{\absangparam_0})}{\Gamma(-n-2\tilde{\absangparam_0}+\frac{1-s}2)} \frac{\Gamma(-2\tilde{\absangparam_0})}{\Gamma(-\tilde{\absangparam_0})}.
    \end{equation}\end{widetext}
    Here we have recovered the unit $r_\text{s}=2M$ for convenience and $\tilde\absangparam$ was estimated by $\tilde\absangparam_0\simeq\tilde\absangparam(\omega_0)$. Since $\tilde{\absangparam_0}$ deviates from an integer, all the Gamma functions here are well-defined. In this work, we focus on the light fermions $Mm<1$ and expand $\omega_0$ to order $(Mm)^4$. 
        
\begin{figure*}[t]
		\centering
		\includegraphics[width=0.98\linewidth]{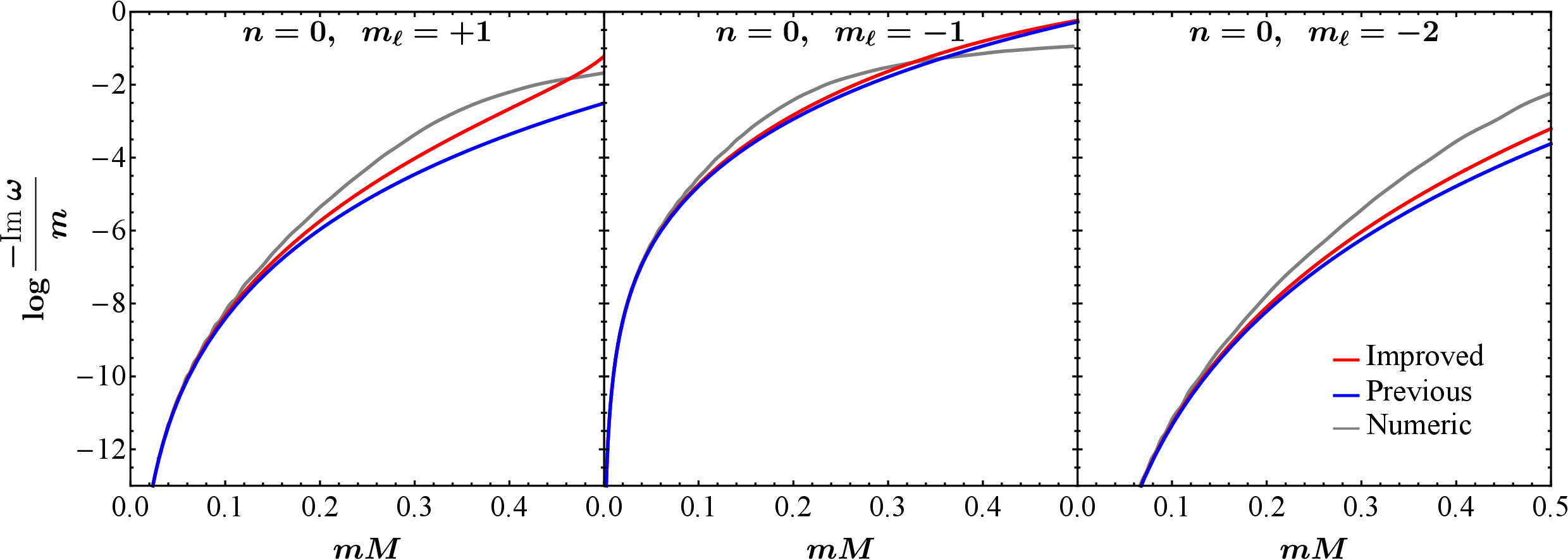}
		\caption{Comparison of improved result Eq.~\eqref{eq_improved_Dirac}, previous result Eq.~\eqref{eq_imagspec_quad_negative}, and numerical calculations collected from Ref.~\cite{Dolan:2015eua} for $n=0$. Left panel for $\angparam=+1$, middle panel for $\angparam=-1$, and right panel for  $\angparam=-2$.}
		\label{fig_comp}
    \end{figure*}

\begin{figure*}[t]
		\centering
		\includegraphics[width=0.98\linewidth]{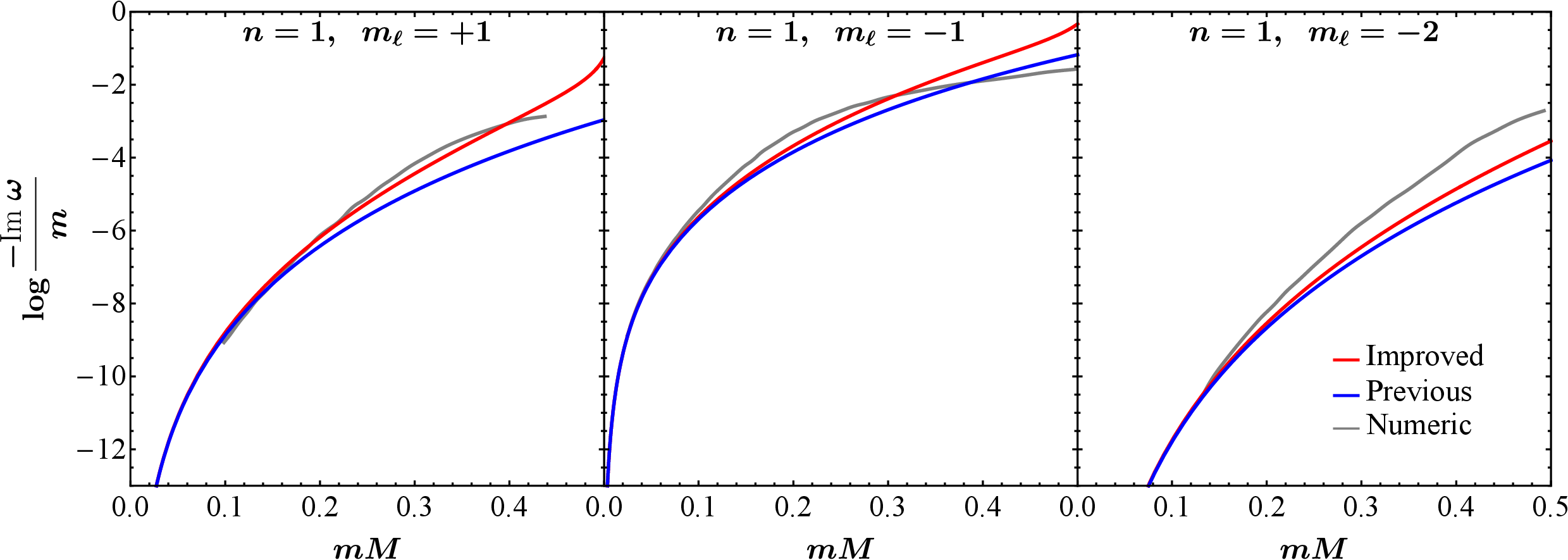}
		\caption{Comparison of improved result Eq.~\eqref{eq_improved_Dirac}, previous result Eq.~\eqref{eq_imagspec_quad_negative}, and numerical calculations collected from Ref.~\cite{Dolan:2015eua} for $n=1$. Left panel for $\angparam=+1$, middle panel for $\angparam=-1$, and right panel for  $\angparam=-2$.}
		\label{fig_comp2}
    \end{figure*}

    The above result provides a well-defined and improved analytic formulation for the spectrum of a spinor fermion moving around the Schwarzschild black hole. The comparisons of analytic results, both outlined in the previous section and presented in this section via the improved formula, with the numerical results are displayed in Figs.~\ref{fig_comp} and \ref{fig_comp2}. The gray curves are the numerical results collected from Ref.~\cite{Dolan:2015eua}. The red curves are the improved results in Eq.~\eqref{eq_improved_Dirac}, and the blue ones are from Eqs.~\eqref{eq_imagspec_quad_positive} and \eqref{eq_imagspec_quad_negative}. The difference between the red curves and the blue ones is due to the nonzero $\epsilon_\absangparam$ in Eq.~\eqref{eq_effective_angparam}, which is
    \begin{equation}
        \begin{aligned}
            \epsilon_\absangparam = \tilde\absangparam -\ell 
            \simeq -\frac{3\pmass^2}{8\absangparam} - {\left(\frac{9}{16\absangparam^2} - \frac{1}{\bar n^2}\right)\frac{m^4}{8\absangparam} } + \dots.
        \end{aligned}
    \end{equation}
    It is seen that $\bar n$ in the case of $\angparam>0$ is larger than that in the case of $\angparam<0$, with the same $n$ and $\absangparam$. This explains why the difference between the red curves and the blue ones for the case $\angparam<0$ is smaller.
	
    \section{Understanding on the well-defined and improved analytic spectrum}
    \label{sec_discussion}

    To justify the well-defined and improved formula in our present considerations, it is crucial to conduct a detailed analysis on the local solutions in regions far from the horizon. It is shown that in the large-$r$ limit, components in Eq.~\eqref{eq_psi_h_Dirac} behave as follows:
    \begin{equation}\label{eq_hierarchy}
		\frac{\tilde\psi^-_f(r)}{\tilde\psi^+_f(r)} \approx \mathcal O(\pmass^{s}).
    \end{equation}
    For a light particle, there is only one component that becomes dominant, which is known as the \textit{large} component. For the case $\angparam=\absangparam$ ($s>0$), the large component represented by $\tilde\psi^+_f$ is proportional to $W_{\tomega,\tilde{\absangparam}+\frac12}$. And for the other case ($s<0$), the roles of $\tilde\psi^-_f$ and $\tilde\psi^+_f$ are interchanged. The large component is now given by $W_{\tomega,\tilde{\absangparam}-\frac12}$. In the nonrelativistic limit, the large component is sufficient to sketch out the structure of the atom. This is precisely what we have outlined in Sec. \ref{sec_review}. 
        
    Near the horizon, the relativistic effect becomes significant, which violates the hierarchy behavior in Eq.~\eqref{eq_hierarchy} as shown in Eq.~\eqref{eq_psif2h}. In this case, both components assume equal significance. For clarity, let us focus on the case $ \angparam=\absangparam$. Within the overlapping region, the wave function has the following form:
    \begin{equation}\label{eq_psif2h_Dirac_full}
        \begin{aligned}
            \tilde\psi_f(r) &\approx (2pr)^{-\tilde{\absangparam}}\frac{2p}{i\beta_-}\frac{\Gamma(2\tilde{\absangparam})}{\Gamma(\tilde{\absangparam}-\tomega)}
            \begin{pmatrix}1\\0\end{pmatrix} \\
            &\ + (2pr)^{\tilde{\absangparam}}\frac{\Gamma(1-2\tilde{\absangparam})}{\Gamma(1-\tilde{\absangparam}-\tomega)}
            \begin{pmatrix}0\\1\end{pmatrix} \\
            &\ + (2pr)^{1+\tilde{\absangparam}}\frac{\tilde{\absangparam}\beta_+}{ip}\frac{\Gamma(-1-2\tilde{\absangparam})}{\Gamma(1-\tilde{\absangparam}-\tomega)} 
            \begin{pmatrix}1\\0\end{pmatrix} \\
            &\ + (2pr)^{1-\tilde{\absangparam}}\frac{\Gamma(-1+2\tilde{\absangparam})}{\Gamma(\tilde{\absangparam}-\tomega)}
            \begin{pmatrix}0\\1\end{pmatrix} ,
        \end{aligned}
    \end{equation}
    where we have used the following identity: 
	\begin{equation}\label{eq_det_indentity}
		\frac{i\tilde{\absangparam}\beta_+}{p(\tilde{\absangparam}+\tomega)}\cdot\frac{i\tilde{\absangparam}\beta_-}{p(\tilde{\absangparam}-\tomega)} = 1,
	\end{equation}
    It indicates that the last two terms in Eq.~\eqref{eq_psif2h_Dirac_full} become subdominant when approaching the horizon. The positive-power terms of the large component are overwhelmed by those of the small component in the overlap region. That is why the solution of the second-order equation~\eqref{eq_psif} cannot be directly matched to the solution in Eq.~\eqref{eq_psih}. In essence, $\tilde\psi^+$ and $\tilde\psi^-$ are of equal importance, which poses challenges for the asymptotic analysis of the second-order equation of motion pertaining to a specific component. 
	
    The analysis of the coupled first-order equation shows that it is the function proportional to
    \begin{equation}
        W_{\tomega,\tilde{\absangparam}+\frac12} + \frac{p(\tilde{\absangparam}+\tomega)}{i\tilde{\absangparam}\beta_+}W_{\tomega,\tilde{\absangparam}-\frac12} ,
    \end{equation}
    which matches with the local solution near the horizon $\psi_h^+$. Nevertheless, the second term $W_{\tilde{\absangparam}-\frac12}$ is absent in the analysis of the second-order equation of motion in the far region. The local solution in the intermediate region indeed serves to compensate for the ignorance. Equation~\eqref{eq_psii} modifies the branches with the positive exponent while keeping the other one. The enhancement factor
        \begin{equation}\label{eq_transition_factor}
    		\frac{2i\pmass}{2\absangparam+1} \approx \frac{i\absangparam\beta_+}{p(\tilde{\absangparam}+\tomega)}
        \end{equation}
    can be interpreted as an implicit jump from the dominant \textit{large} component to the \textit{small} component.

    \section{Summary}
    \label{sec_summary}
    We have revisited the fermionic quasibound state around a Schwarzschild black hole and obtained an \textit{improved} analytic formula by matching local solutions within two overlapping regions. A careful analysis of the coupled first-order radial equations of motion reveals that the dominant large component in the far region fails to accurately represent the wave function near the horizon, even when maintaining a significant distance from it. Consequently, one cannot directly match the solution with another approximated solution near the horizon. It is necessary to introduce one more solution in an additional intermediate region serving as a bridge~\cite{Ternov:1980st}. On the other hand, while the two components in the coupled equations are addressed in the region of large $r$, their behaviors are automatically adjusted to approach the horizon in the right way and can be matched with the solution near the horizon directly.
    
    To improve the analytical result, a correction to the quantum number $\tilde\absangparam=\absangparam+\epsilon_\absangparam$ has been taken into consideration. The inclusion of $\epsilon_\ell$ can avoid the ill-defined Gamma functions and introduce some visible improvements to $\omega$. The imaginary part of the result agrees with the solutions~\eqref{eq_imagspec_quad_positive} and~\eqref{eq_imagspec_quad_negative} but is enhanced for the value $m M$ within 0.1-0.5. The improvement can be seen for the case of $\angparam=1$. Since $\epsilon_\ell\sim m^2/\ell$ and is smaller for $m_\ell=-\ell$ than that for $m_\ell=\ell$, the improvement decreases when $m_\absangparam$ is negative or $n$ and $\absangparam$ get larger. Furthermore, the improved formula provides a well-defined compact form, which also confirms the proposed regularization~\eqref{eq_factor2}. The nonzero $\epsilon_\ell$ also gives a correction to the real part of the frequency, which is small and at the order of $m^4$.
    
    \begin{acknowledgments}
        This work was supported in part by the National Science Foundation of China (NSFC) under Grants No. 12147103 (special fund to the center for quanta-to-cosmos theoretical physics), No. 11821505, No. 12447105, and by the National Key Research and Development Program of China under Grant No.2020YFC2201501, the Strategic Priority Research Program of the Chinese Academy of Sciences under Grant No. XDB23030100.
    \end{acknowledgments}

\appendix

    \section{A BRIEF INTRODUCTION TO GQFT}\label{app_GQFT}
To outline the framework of GQFT, we start with the action of a free Dirac fermion in quantum field theory,
\begin{equation} \label{actionDF}
\begin{aligned}
    S_{DF} = \int d^4x\,  \frac{1}{2} \left( \bar{\psi}(x) \gamma^{a} \delta_{a}^{\;\;\mu} i \partial_{\mu} \psi(x) + H.c. \right) & \\
    - m\, \bar{\psi}(x) \psi(x) &
\end{aligned}
\end{equation}
where $\psi(x)$ is a complex Dirac spinor field with mass $m$, $\gamma^{a}$ ($a=0,1,2,3$) are the Dirac $\gamma$-matrices, and $\delta_{a}^{\;\;\mu} $ denotes the Kronecker symbol. Here, the Greek alphabet ($\mu,\nu = 0,1,2,3$) and the Latin alphabet ($a,b=0,1,2,3$) are used to distinguish the coordinate vectors from the spin vectors, which are raised and lowered by the constant metric tensors $\eta^{\mu\nu}(\eta_{\mu\nu})=(1,-1,-1,-1)$ and $\eta^{ab}(\eta_{ab})= (1,-1,-1,-1)$. 

The above action is known to exhibit an associated symmetry,
\begin{align}
G_S = \mathrm{PO(1,3)}\widetilde{\Join} \mathrm{SP(1,3)} = \mathrm{P^{1,3}}\ltimes \mathrm{SO(1,3)}\widetilde{\Join} \mathrm{SP(1,3)} ,
\end{align}
where PO(1,3) $\equiv \mathrm{P^{1,3}}\ltimes$SO(1,3) represents the Poincar\'e symmetry group (or inhomogeneous Lorentz symmetry). The symbol ``$\widetilde{\Join}$" denotes the associated symmetry, indicating that the transformation of the spin symmetry SP(1,3), in spinor representation of the Dirac fermion, must align with that of the isomorphic Lorentz symmetry SO(1,3) in Minkowski spacetime. This associated symmetry forms the foundation for deriving the Dirac equation of the electron. Here, P$^{1,3}$ corresponds to the translational symmetry in Minkowski spacetime. 

The gauge invariance principle states that the laws of nature should be independent of the choice of local field configurations. This implies that internal symmetries must be localized as gauge symmetries to describe fundamental interactions. Consequently, the internal spin symmetry SP(1,3) for the Dirac fermion $\psi(x)$ must be a local gauge symmetry, while the Poincar\'e group symmetry PO(1,3) of coordinates retains its global symmetry. 

To preserve the local spin gauge symmetry SP(1,3) alongside the global Lorentz symmetry SO(1,3), we should introduce a spin gauge field $\cA_{\mu}(x)$ to ensure the spin gauge symmetry SP(1,3), and a bicovariant vector field $\hat{\chi}_{a}^{\; \mu}(x)$ to replace the Kronecker delta symbol $\delta_{a}^{\;\;\mu}$. This extension modifies the ordinary derivative $\partial_{\mu}$, associated with the delta symbol $\delta_{a}^{\;\;\mu}$ in Eq.(\ref{actionDF}), into a covariant derivative:  
\begin{equation}
    \begin{aligned}
        \delta_{a}^{\;\;\mu} i\partial_{\mu} \to i\hat{\chi}_{a}^{\; \, \mu} \mathcal{D}_{\mu} &\equiv \hat{\chi}_{a}^{\; \, \mu}( i\partial_{\mu} + g_s\cA_{\mu} ) \\
        &\equiv  i\eth_{a} + g_s\hcA_{a} \\
        &\equiv  i\hcD_{a} , 
    \end{aligned}
\end{equation}
where
\begin{align}
\cA_{\mu} & \equiv \cA_{\mu}^{bc} \frac{1}{2} \Sigma_{bc} ,\quad \hcA_{a} \equiv \hat{\chi}_{a}^{\; \, \mu} \cA_{\mu} .
\end{align}
Here, $\Sigma_{bc}= \dfrac{i}{4}[\gamma_b, \gamma_c]$ are the generators of SP(1,3). 

The spin-related intrinsic derivative operator $\eth_{a}$ defines a corresponding intrinsic displacement vector $\dbar\zeta^{a}$. Their relationship to the ordinary coordinate derivative operator $\partial_{\mu}\equiv {\partial}/{\partial x^{\mu}}$ and displacement vector $dx^{\mu}$ is mediated by the dual bicovariant fields  $\hat{\chi}_{a}^{\; \mu}(x)$ and $\chi_{\mu}^{\; a}(x)$:
\begin{align}
 \eth_{a}  \equiv  \hat{\chi}_{a}^{\; \mu}(x)\p_{\mu}, \quad \dbar\zeta^{a} \equiv  \chi_{\mu}^{\; a}(x) dx^{\mu}, 
\end{align}
where $\hat{\chi}_{a}^{\; \mu}(x)$ and $\chi_{\mu}^{\; a}(x)$ satisfy the following dual conditions:
\begin{align}
\chi_{\mu}^{\; a}(x) \hat{\chi}_{b}^{\; \nu}(x)  \eta_{a}^{\; b} = \eta_{\mu}^{\; \nu} , \quad 
\hat{\chi}_{b}^{\; \nu}(x)  \chi_{\mu}^{\; a}(x) \eta_{\nu}^{\; \mu} = \eta_{b}^{\; a} .
\end{align}

The dual vectors $\hat{\chi}_{a}^{\; \mu}(x)$ and $\chi_{\mu}^{\; a}(x)$ transform as bicovariant vector fields under both the spin gauge symmetry SP(1,3) and the global Lorentz symmetry SO(1,3):
\begin{equation} \label{GT}
\begin{aligned}
\chi_{\mu}^{\; a}(x) \to \chi_{\mu}^{'\; a}(x) &=  \Lambda^{a}_{\; b}(x)\chi_{\mu}^{\; b}(x), \\
\hat{\chi}_{a}^{\; \mu}(x) \to \hat{\chi}_{a}^{'\; \mu}(x) &=  \Lambda_{a}^{\; b}(x) \hat{\chi}_{b}^{\; \mu}(x), \\
\chi_{\mu}^{\; a}(x) \to \chi_{\mu}^{'\; a}(x') &=  L_{\mu}^{\; \nu}\, \chi_{\nu}^{\; a}(x), \\
\hat{\chi}_{a}^{\; \mu}(x) \to \hat{\chi}_{a}^{'\; \mu}(x') &=  L^{\mu}_{\; \nu}\, \hat{\chi}_{a}^{\; \nu}(x) ,\\ 
x^{' \mu} &= L^{\mu}_{\; \nu}\, x^{\nu} ,
\end{aligned}
\end{equation}
with $\Lambda^{a}_{\; b}(x) \in$ SP(1,3) $\cong$ SO(1,3), and $L^{\mu}_{\; \nu} \in$SO(1,3).

In the external spacetime of coordinates, the derivative operator $\p_{\mu}$ and displacement vector $dx^{\mu}$ form dual bases $\{\p_{\mu}\}\equiv \{\partial/\partial x^{\mu}\}$ and $\{dx^{\mu}\}$, spanning the tangent spacetime $T_{4}$ and cotangent spacetime $T^{*}_{4}$, respectively. Analogously, the intrinsic derivative operator $\eth_a$ and the displacement vector $\dbar \zeta^{a}$ define dual bases $\{\eth_a\}$ and $\{\dbar \zeta^{a}\}$, generating the intrinsic spacetime and its dual. We may denote the spin-related intrinsic spacetime and its dual as $G_{4}$ and $G^{*}_{4}$ respectively.

The intrinsic basis $\{\eth_a \}$ satisfies the commutation relation:
\begin{align} \label{NCR}
[ \eth_c ,\; \eth_d] = \hat{\mathsf{F}}_{cd}^a \eth_a ,\quad \hat{\mathsf{F}}_{cd}^a \equiv  - \hat{\chi}_{c}^{\; \mu} \hat{\chi}_{d}^{\; \nu} \mathsf{F}_{\mu\nu}^{a},
\end{align}
where $\hat{\mathsf{F}}_{cd}^a$ acts as group structure factor, and $\mathsf{F}_{\mu\nu}^{a}$ represents a gauge-type field strength, explicitly given by
\begin{align}  \label{GFS}
\mathsf{F}_{\mu\nu}^{a} \equiv \p_{\mu}\chi_{\nu}^{\; a}(x) - \p_{\nu}\chi_{\mu}^{\; a}(x) .
\end{align}
This confirms that $\chi_{\mu}^{\; a}(x)$ behaves as a gauge-type bicovariant vector field defined in Minkowski spacetime and valued in spin-related intrinsic spacetime. 

The spin-related gauge-type bicovariant vector field $\chi_{\mu}^{\; a}(x)$ naturally emerges to describe the gravitational interaction associated with the spin gauge symmetry of the spinor field as a basic constituent of matter. We therefore designate it as the spin-related gravigauge field, with $\mathsf{F}_{\mu\nu}^{a}$ representing its gravigauge field strength. The spin-related intrinsic spacetime $G_{4}$ and its dual $G^{*}_{4}$, characterized by $\chi_{\mu}^{\; a}(x)$, are collectively termed gravigauge spacetime. 

In this framework, the global flat Minkowski spacetime serves as the base spacetime and the gravigauge spacetime constitutes the fiber. The intrinsic derivative operator $\eth_{a}$ and the displacement vector $\dbar \zeta^{a}$ are called the gravigauge derivative and the gravigauge displacement, respectively. The gravigauge spacetime can alternatively be characterized by a spin connection $\mOm_{c}^{ab}$: 
\begin{equation} \label{SGGF}
\begin{aligned}
\mOm_{c}^{ab} &\equiv \hat{\chi}_{c}^{\; \mu} \mOm_{\mu}^{ab} ,\\
\mOm_{\mu}^{ab}(x) &= \frac{1}{2}\left( \hat{\chi}^{a\nu} \mathsf{F}_{\mu\nu}^{b} - \hat{\chi}^{b\nu} \mathsf{F}_{\mu\nu}^{a} -  \hat{\chi}^{a\rho}  \hat{\chi}^{b\sigma}  \mathsf{F}_{\rho\sigma}^{c} \chi_{\mu c } \right) .
\end{aligned}
\end{equation}
This spin connection, determined entirely by the gravigauge field, is referred to as the spin gravigauge field.

The group structure factor $\hsF_{cd}^a$ and the spin gravigauge field $\mOm_{c}^{ab}$ are fundamentally related through:
\begin{equation}
\begin{aligned}
\hsF_{cd}^{a}  &= \hmOm_{cd}^{a}  - \hmOm_{dc}^{a}, \\
\hmOm_{c}^{ab} &\equiv \etach_{ca'}^{[ab] c'd'} \hsF_{c'd'}^{a'}  ,\\
\etach_{ca'}^{[ab] c'd'} &\equiv \frac{1}{2}  (\eta^{ac'}\eta_{a'}^{\, b} - \eta^{bc'}\eta_{a'}^{\, a}) \eta_{c}^{\, d'} \\
&\qquad+ \frac{1}{4} (\eta^{ac'}\eta^{bd'} - \eta^{bc'}\eta^{ad'} ) \eta_{ca'} .
\end{aligned}
\end{equation}

In the gravigauge spacetime, the gauge covariant field strength $\hcF_{cd}$ takes the form: 
\begin{equation}
\begin{aligned}
    \hcF_{cd} &\equiv  \mathcal{D}_{c}\hcA_{d} - \mathcal{D}_{d}\hcA_{c} - i g_s [\hcA_{c}, \hcA_{d}] \\
    &\equiv \cF_{cd} + \hsF_{cd}^{a} \hcA_{a}, \\
    \mathcal{D}_{c}\hcA_{d} &\equiv \eth_{c} \hcA_{d} + \hmOm_{cd}^{a} \hcA_{a} , \\
    \hsF_{cd}^{a} \hcA_{a} &\equiv (\hmOm_{cd}^{a} - \hmOm_{dc}^{a}) \hcA_{a}, \\
    \cF_{cd} &\equiv \eth_{c}\hcA_{d} - \eth_{d}\hcA_{c} - i g_s [\hcA_{c}, \hcA_{d}] , 
\end{aligned}
\end{equation}
where the term $\hsF_{cd}^{a}\hcA_{a}$ arises from gravitational effects due to the noncommutative nature of the gravigauge spacetime.

Based on the internal spin gauge symmetry SP(1,3) analyzed above, we can extend the action for the free Dirac fermion given in Eq. (\ref{actionDF}) to the following gravitational action of the Dirac fermion in spin-related gravigauge spacetime:
\begin{equation} \label{actionGDF1}
\begin{aligned}
    \cS_{\mathrm{GDF}} =&\ \int [\dbar \zeta^{c}]\, \frac{1}{2} \left( \bar{\psi}(x) \gamma^{c} i \hcD_{c} \psi(x) + H.c. \right) \\
    &\ - m\, \bar{\psi}(x) \psi(x)  - \frac{1}{2} \eta^{c c'}\eta^{d d'} \Tr \hcF_{cd}\hcF_{c'd'} \\
    &\ + M_{\cA}^2 \eta^{cd}  \Tr (\hcA_{c} -  \hmOm_{c}/g_s)( \hcA_{d}-\hmOm_{d}/g_s) \\
    &\ + \frac{1}{4} \bM_{\kappa}^2 \etat^{cdc'd'}_{aa'} \hsF_{cd}^{a}\hsF_{c'd'}^{a'},
\end{aligned}
\end{equation}
where $\bM_{\kappa}$ is a fundamental mass scale and $M_{\cA}$ denotes the mass of the spin gauge field. 

The constant tensor $\etat^{cdc'd'}_{aa'}$ is defined as
\begin{equation} \label{CTensor1}
\begin{aligned}
    \tilde{\eta}^{cd c'd'}_{a a'} &\equiv \eta^{c c'} \eta^{d d'} \eta_{a a'}  
+  \eta^{c c'} ( \eta_{a'}^{d} \eta_{a}^{d'}  -  2\eta_{a}^{d} \eta_{a'}^{d'}  ) \\ 
&\quad\ +  \eta^{d d'} ( \eta_{a'}^{c} \eta_{a}^{c'} -2 \eta_{a}^{c} \eta_{a'}^{c'} ) . 
\end{aligned}
\end{equation}
which possesses a special structure to ensure the spin gauge invariance for the action.

The gravigauge field $\chi_{\mu}^{\;\; a}$, an invertible bicovariant vector field, behaves as a Goldstone boson. This field facilitates the projection of vectors and tensors between gravigauge spacetime and Minkowski spacetime. Consequently, the action in gravigauge spacetime [Eq.~\eqref{actionGDF1}] can be reformulated in Minkowski spacetime as follows:
\begin{equation} \label{actionGDF2}
\begin{aligned}
    \cS_{\mathrm{GDF}}  & \equiv \int [d x]\chi \cL_{\mathrm{GDF}} \\
    &= \int d^4x\,  \chi \bigg\{ \frac{1}{2} \left( \bar{\psi}(x) \gamma^{a} \hat{\chi}_{a}^{\; \mu} i\mathcal{D}_{\mu} \psi(x) + H.c. \right)   \\
    &\ - m\, \bar{\psi}(x) \psi(x) - \frac{1}{4} \hat{\chi}^{\mu\mu'}\hat{\chi}^{\nu \nu'} \cF_{\mu\nu}^{ab}\cF_{\mu'\nu' ab}\\
    &\ + \frac{1}{2} M_{\cA}^2 \hat{\chi}^{\mu\nu} (\cA_{\mu}^{ab} -  \mOm_{\mu}^{ab}/g_s)( \cA_{\nu ab} -  \mOm_{\nu ab}/g_s)   \\
    &\  + \frac{1}{4} \bM_{\kappa}^2 \tilde{\chi}^{\mu\nu\mu'\nu'}_{aa'} \mathsf{F}_{\mu\nu}^{a}\mathsf{F}_{\mu'\nu'}^{a'}\bigg\},
\end{aligned}
\end{equation}
with the tensors defined as follows,
\begin{align}
& \tilde{\chi}_{aa'}^{\mu\nu \mu'\nu'} \equiv \hat{\chi}_{c}^{\;\, \mu}\hat{\chi}_{d}^{\;\, \nu} \hat{\chi}_{c'}^{\;\, \mu'} \hat{\chi}_{d'}^{\;\, \nu'}  \etat^{c d c' d'}_{a a'} , \nn \\
&  \hat{\chi}^{\mu\nu} \equiv \hat{\chi}_{a}^{\; \mu} \hat{\chi}_{b}^{\; \nu} \eta^{ab} , \quad \chi_{\mu\nu} \equiv \chi^{\;a}_{\mu} \chi^{\; b}_{\nu} \eta_{ab} , \nn \\
& \chi = \det \chi_{\mu}^{\; a} = \sqrt{-\det \chi_{\mu\nu}}, 
\end{align} 
where the tensor field $\chi_{\mu\nu}$ represents the gravimetric field.

We can verify the following relation for the spin gauge invariant masslike term:
\begin{equation} \label{MSG}
\begin{aligned}
    \frac{1}{2} \hat{\chi}^{\mu\nu}  ( \cA_{\mu}^{ab}-\mOm_{\mu}^{ab} /g_s)  ( \cA_{\nu ab}-\mOm_{\nu ab}/g_s) &\\
    = \frac{1}{4 g_s^2} \bchi^{\mu\nu \mu' \nu'}\cG_{\mu\nu}^{a}\cG_{\mu'\nu' a} &,  
\end{aligned}
\end{equation}
where the spin gauge covariant field strength is defined as
\begin{align}\label{GCFS}
&  \cG_{\mu\nu}^{a} \equiv  \p_{\mu} \chi_{\nu}^{\; a}  - \p_{\nu} \chi_{\mu}^{\; a} + g_s(\cA_{\mu b}^{a} \chi_{\nu}^{\; b}  - \cA_{\nu b}^{a} \chi_{\mu}^{\; b} ),
\end{align}
with the tensor structure given by,
\begin{align}
& \chib_{aa'}^{\mu\nu \mu'\nu'} \equiv \hat{\chi}_{c}^{\;\, \mu}\hat{\chi}_{d}^{\;\, \nu} \hat{\chi}_{c'}^{\;\, \mu'} \hat{\chi}_{d'}^{\;\, \nu'}  \etab^{c d c' d'}_{a a'} , \nn \\
& \etab^{c d c' d'}_{a a'} \equiv \frac{3}{2}  \eta^{c c'} \eta^{d d'} \eta_{a a'}  
-  \frac{1}{2} ( \eta^{c c'} \eta_{a'}^{d} \eta_{a}^{d'}  +  \eta^{d d'} \eta_{a'}^{c} \eta_{a}^{c'} ) .
\end{align}
This demonstrates that both the spin gravigauge field $\mOm_{\mu}^{ab}$ and the spin gauge field $g_s\cA_{\mu}^{ab}$ transform identically under the spin gauge symmetry SP(1,3). 

The purely gravitational interactions characterized by the gravigauge field strength $\mathsf{F}_{\mu\nu}^{a}$ in the above action are equivalent to the Einstein-Hilbert action up to a total derivative term,
\begin{align} \label{GGI}
 \frac{1}{4} \chi\, \tchi_{aa'}^{\mu\nu \mu'\nu'} \mathsf{F}_{\mu\nu}^{a} \mathsf{F}_{\mu'\nu'}^{a'} \equiv \chi\, R
- 2 \p_{\mu} (\chi \hat{\chi}^{\mu\rho} \hat{\chi}_{a}^{\;\sigma} \mathsf{F}_{\rho\sigma}^{a} ) , 
\end{align}
where $R$ denotes the Ricci curvature scalar, defined through the following identities:
\begin{align}  \label{GGGI}
 & R \equiv  \hat{\chi}_{b}^{\; \mu} \hat{\chi}_{a}^{\; \nu} R_{\mu\nu}^{ab}  \equiv \hat{\chi}^{\mu\sigma} \hat{\chi}^{\nu\rho} R_{\mu\nu\rho\sigma} \equiv \hat{\chi}^{\mu\sigma} R_{\mu\sigma} , 
\end{align}
 with the field strength $R_{\mu\nu}^{ab}$ of spin gravigauge field and Riemann curvature tensor $R_{\mu\nu\rho\sigma}\equiv \chi_{\rho\lambda}R_{\mu\nu\sigma}^{\;\lambda}$ given by
\begin{align} 
 & R_{\mu\nu}^{ab} = \p_{\mu}\mOm_{\nu}^{ab} - \p_{\nu}\mOm_{\mu}^{ab} + \mOm_{\mu c}^{a} \mOm_{\nu}^{cb} - \mOm_{\nu c}^{a} \mOm_{\mu}^{cb} , \nn \\
 & R_{\mu\nu\sigma}^{\;\rho}(x)  = \p_{\mu} \Gamma_{\nu\sigma}^{\rho} - \p_{\nu} \Gamma_{\mu\sigma}^{\rho}  + \Gamma_{\mu\lambda}^{\rho} \Gamma_{\nu\sigma}^{\lambda}  - \Gamma_{\nu\lambda}^{\rho} \Gamma_{\mu\sigma}^{\lambda} .
\end{align}
The affine connection (or Christoffel symbol) $\Gamma_{\mu\sigma}^{\rho}(x)$ relates to the spin gravigauge field $\mOm_{\mu}^{ab}$ through
\begin{align}  \label{RMC}
& \Gamma_{\mu\sigma}^{\rho}(x)  \equiv  \hat{\chi}_{a}^{\;\; \rho} \p_{\mu} \chi_{\sigma}^{\;\; a} 
+  \hat{\chi}_{a}^{\;\; \rho}   \mOm_{\mu b}^{a} \chi_{\sigma}^{\;\;b} , \nn \\
& \quad \quad \quad = \frac{1}{2}\hat{\chi}^{\rho\lambda} (\p_{\mu} \chi_{\lambda\sigma} + \p_{\sigma} \chi_{\lambda\mu} - \p_{\lambda}\chi_{\mu\sigma} ).
\end{align}

Using these relations and identities, we can reformulate the GQFT action from Eq. (\ref{actionGDF2}) as follows:
\begin{equation} \label{actionGDF3}
\begin{aligned}
    \cS_{\mathrm{GDF}} &= \int d^4x\,  \chi \bigg\{ \frac{1}{2} \left( \bar{\psi}(x) \gamma^{a} \hat{\chi}_{a}^{\; \mu} i\mathcal{D}_{\mu} \psi(x) + H.c. \right) \\ 
    &\ - m\, \bar{\psi}(x) \psi(x) - \frac{1}{4} \hat{\chi}^{\mu\mu'}\hat{\chi}^{\nu \nu'} \cF_{\mu\nu}^{ab}\cF_{\mu'\nu' ab} \\
    &\ + \frac{1}{4}\frac{M_{\cA}^2}{g_s^2} \bchi^{\mu\nu \mu' \nu'}\cG_{\mu\nu}^{a}\cG_{\mu'\nu' a}  + \bM_{\kappa}^2 R  \bigg\} .
\end{aligned}
\end{equation}

The actions defined in Eqs. (\ref{actionGDF2}) and (\ref{actionGDF3}) exhibit a hidden general linear group symmetry GL(1,3,R) in coordinate spacetime. This becomes evident when examining the fundamental action in Eq. (\ref{actionGDF1}), which is formulated in gravigauge spacetime and should remain invariant under arbitrary coordinate transformations. Consequently, the action in Eq. (\ref{actionGDF3}) acquires an enhanced joint symmetry:
\begin{align} \label{EJS}
G_S = \mbox{GL(}1,3,\mbox{R)} \Join \mathrm{SP(1, 3)}.
\end{align}
This demonstrates how the global Poincar\'e symmetry PO(1,3) underlying GQFT naturally extends to a local GL(1,3,R) symmetry when constructing actions in gravigauge spacetime through the gauge invariance principle.

The least action principle enables one to derive equations of motion for all fields. For the Dirac fermion field $\psi(x)$,we obtain
\begin{align}  \label{EoMD}
\gamma^{a} \hat{\chi}_{a}^{\; \mu} i( \mathcal{D}_{\mu}  - \mV_{\mu}) \psi(x) - m \psi(x) = 0, 
\end{align}
where the induced vector gauge field $\mV_{\mu}$ is defined as
\begin{align} 
\mV_{\mu} & \equiv  \frac{1}{2} \chi \, \hat{\chi}_{b}^{\;\; \nu}\mathcal{D}_{\nu}(\hat{\chi} \chi_{\mu}^{\;\; b}), 
\end{align}
with the covariant derivative operation
\begin{align} 
 \mathcal{D}_{\nu}(\hat{\chi}\chi_{\mu}^{\;\;a}) = \p_{\nu} (\hat{\chi}\chi_{\mu}^{\;\; a}) + \hat{\chi} g_s\cA_{\nu\, b}^{a}  \chi_{\mu}^{\;\;b} .
\end{align}

The equations of motion for the gravigauge field $\chi_{\mu}^{\;\; a}(x)$ are obtained as
\begin{align}   \label{GaugeGE}
 \p_{\mu} \whsF^{\mu\nu }_{a}  = \whmJ_{a}^{\; \nu}   ,\qquad \p_{\nu} \whmJ_{a}^{\; \nu} = 0, 
\end{align}
where the field strength $\whsF^{\mu\nu }_{a}$ is defined by, 
\begin{equation}
\begin{aligned}
    \whsF^{\mu\nu }_{a} &\equiv \chi  \tchi^{[\mu\nu]\rho\sigma}_{a b}  \mathsf{F}_{\rho\sigma }^{b}  , \\
    \tchi^{[\mu\nu]\mu'\nu'}_{a a'} &\equiv \frac{1}{2}( \tchi^{\mu\nu\mu'\nu'}_{a a'} - \tchi^{\nu\mu\mu'\nu'}_{a a'} ) ,
\end{aligned}
\end{equation}
and the current $\whmJ_{a}^{\; \nu}$ (whose conservation follows from the antisymmetry of $\whsF^{\mu\nu }_{a}$) takes the explicit form:
\begin{align} 
\whmJ_{a}^{\; \nu}  & \equiv  16\pi G_{\kappa} \chi [  \hat{\chi}_{a}^{\; \nu}  \cL_{\mathrm{GDF}}  - \hat{\chi}_{b}^{\; \nu}  \hat{\chi}_{a}^{\;\rho} \frac{1}{2} ( \bar{\psi}\gamma^{b} i\mathcal{D}_{\rho} \psi   + H.c.  ) \nn \\
&\quad +   \hat{\chi}_{a}^{\; \; \rho}  \hat{\chi}^{\nu\sigma} \hat{\chi}^{\rho'\sigma'}  \cF_{\rho\rho'}^{bc} \bscF_{\sigma\sigma' bc}   ]  -  \hat{\chi}_{a}^{\; \rho}  \mathsf{F}_{\rho\sigma}^{b}  \whsF^{\nu\sigma}_{b}\nn \\
 &\quad -   \gamma_{W} \mathcal{D}_{\rho}\widehat{\cG}_{a}^{\rho\nu} - \gamma_{W}\hat{\chi}_{a}^{\; \rho}  \cG_{\rho\sigma}^{b} \widehat{\cG}^{\nu\sigma}_{b} ,
\end{align}
with the following fundamental definitions: 
\begin{equation} \label{GGFS}
\begin{aligned}
    16\pi G_{\kappa} &\equiv \frac{1}{\bM_{\kappa}^2}, \quad 
    \gamma_W \equiv  \frac{M_{\cA}^2}{g_s^2\bM_{\kappa}^2}  ,  \\
    \whcG_a^{\mu\nu} &\equiv \chi\, \bchi^{[\mu\nu]\mu'\nu'}_{a a'}  \cG_{\mu'\nu' }^{a'}, \\ 
    \bchi^{[\mu\nu]\mu'\nu'}_{a a'} &\equiv \frac{1}{2}( \bchi^{\mu\nu\mu'\nu'}_{a a'} - \bchi^{\nu\mu\mu'\nu'}_{a a'} ) .
\end{aligned}
\end{equation}

The gauge-type gravitational equation [Eq~\eqref{GaugeGE}] provides a complete description of gravidynamics within the GQFT framework, exhibiting a formal analogy with Maxwell's equations in electromagnetism.

To obtain the geometric formulation analogous to Einstein's equations, we project Eq. (\ref{GaugeGE}) onto the coordinate spacetime using the gravigauge field $\chi_{\mu}^{\; a}$. This yields the general geometric-type gravitational equation:
 \begin{align} \label{GGE}
 R_{\mu\nu} - \frac{1}{2} \chi_{\mu\nu} R  +  \gamma_W \cG_{\mu\nu} = 8\pi G_{\kappa} \mT_{\mu\nu} ,
 \end{align}
where the tensors $\mT_{\mu\nu}$ and $\cG_{\mu\nu}$ are given by
\begin{equation} \label{CC}
\begin{aligned}
    \mT_{\mu\nu} &=  \frac{1}{2} ( \chi_{\mu\nu} \hat{\chi}_{a}^{\; \rho}   - \eta_{\mu}^{\; \rho}  \chi_{\nu a} ) ( \bar{\psi} \gamma^{a} i\mathcal{D}_{\rho} \psi + H.c.)\\
    &\quad - \chi_{\mu\nu} m \bar{\psi}\psi  \\
    &\quad\ + ( \eta_{\mu}^{\; \rho}\eta_{\nu}^{\; \sigma} - \frac{1}{4} \chi_{\mu\nu}  \hat{\chi}^{\rho\sigma} )\hat{\chi}^{\rho'\sigma'} \cF_{\rho\rho'}^{ab} \cF_{\sigma\sigma' ab}, \\
\cG_{\mu\nu} &= \frac{1}{2} \hat{\chi}\, \chi_{\mu}^{\;\, a}\mathcal{D}_{\rho}(\widehat{\cG}_{a}^{\rho\sigma})\chi_{\sigma\nu}\\
&\quad +  \frac{1}{2}  (\eta_{\mu}^{\; \rho}\chi_{\lambda\nu } -  \frac{1}{4} \chi_{\mu\nu}\eta_{\lambda}^{\; \rho} ) \hat{\chi} \cG_{\rho\sigma}^{a} \widehat{\cG}^{\lambda\sigma}_{a}. 
\end{aligned}
\end{equation}

In contrast to GR, the spin gauge-invariant tensors $\mT_{\mu\nu}$ and $\cG_{\mu\nu}$ in GQFT are generically asymmetric ($\mT_{\mu\nu} \neq \mT_{\nu\mu}$, $\cG_{\mu\nu} \neq \cG_{\nu\mu}$). This allows decomposition into symmetric and antisymmetric components:
 \begin{equation} \label{GGE2}
 \begin{aligned}
     R_{\mu\nu} -  \frac{1}{2}\chi_{\mu\nu} R  + \gamma_W \cG_{(\mu\nu)} &=  8\pi G_{\kappa} \mT_{(\mu\nu)} , \\
     \gamma_W \cG_{[\mu\nu]}  &=  8\pi G_{\kappa} \mT_{[\mu\nu]} , 
 \end{aligned}
\end{equation}
where we define the symmetric and antisymmetric parts through:
\begin{equation}
\begin{aligned}
     \mT_{\mu\nu}  & \equiv \mT_{(\mu\nu)} + \mT_{[\mu\nu]}, & 
     \cG_{\mu\nu}   &\equiv \cG_{(\mu\nu)} + \cG_{[\mu\nu]},  \\
    \mT_{(\mu\nu)}  & \equiv  \frac{1}{2} (\mT_{\mu\nu} + \mT_{\nu\mu} ), & \mT_{[\mu\nu]}  &\equiv  \frac{1}{2} (\mT_{\mu\nu} - \mT_{\nu\mu} ),  \\
    \cG_{(\mu\nu)}  & \equiv  \frac{1}{2} (\cG_{\mu\nu} + \cG_{\nu\mu} ), & \cG_{[\mu\nu]}  &\equiv  \frac{1}{2} (\cG_{\mu\nu} - \cG_{\nu\mu} ).
\end{aligned}
\end{equation}

The tensor term $\cG_{\mu\nu}$ introduces a fundamental extension beyond GR. The symmetric component of the field equations yields a generalized Einstein equation, while the antisymmetric component provides an additional independent equation. This structure arises because the spin-related gravigauge field serves as the fundamental gravitational field and the spin gauge field as the basic gauge field, both respecting the spin gauge symmetry SP(1,3). The masslike term for the spin gauge field with the relation presented in Eq.~\eqref{MSG} plays a crucial role in ensuring theoretical consistency when the gravigauge field, serving as the fundamental gravitational field, couples to Dirac spinor fields. This framework leads to several remarkable features of GQFT that have been investigated in recent works~\cite{Wu:2024mul, Wu:2025abi}, most notably: the zero energy-momentum tensor theorem with the cancellation law in GQFT, and a unified theoretical framework that reconciles the standard model of particle physics (formulated in quantum field theory) and the standard cosmological model (based on GR).


\end{document}